\documentclass[conference]{IEEEtran-v17}

\pdfoutput=1

\usepackage{xspace}
\usepackage{url}
\usepackage[cmex10]{amsmath}
\usepackage{bbm}
\usepackage{graphicx}

\usepackage{ifthen}

\newboolean{conf}
\setboolean{conf}{false} 

\usepackage{fancyref} 
\usepackage{amsmath}
\usepackage{amssymb}
\usepackage{dsfont}
\usepackage{footnote}
\usepackage[nosort]{cite}
\usepackage[british]{babel}

\usepackage{color}
\definecolor{links}{rgb}{0.7,0,0}   
\definecolor{urls}{rgb}{0,0,0.8}    
\definecolor{cites}{rgb}{0,0,0.8}   

\usepackage[colorlinks,hyperindex,linkcolor=links,citecolor=cites,urlcolor=urls]{hyperref}

\usepackage{bm}
\usepackage{upgreek}
\usepackage{amssymb}
\usepackage{amsfonts}
\usepackage{mathrsfs}
\usepackage{xspace}

\makeatletter
\def\amsbb{\use@mathgroup \M@U \symAMSb}
\makeatother

\newcommand{\lefto}{\mathopen{}\left}
\newcommand{\safemath}[2]{\newcommand{#1}{\ensuremath{#2}\xspace}}
\safemath{\opE}{\amsbb{E}}
\newcommand{\Ex}[2]{\ensuremath{\amsbb{E}_{#1}\mathopen{}\left[#2\right]}} 	
\safemath{\prob}{\amsbb{P}}
\safemath{\bigO}{\mathcal{O}}
\safemath{\littleo}{\mathit{o}}

\safemath{\extendreal}{\overline{\realset}}

\newcommand{\tp}[1]{\ensuremath{#1^{\mathrm{T}}}} 		

\newcommand{\fnorm}[1]{\ensuremath{\left\|#1\right\|_{\mathsf{F}}}}

\newtheorem{thm}{Theorem}
\newtheorem{lemma}[thm]{Lemma}

\newtheorem{rem}{Remark}

\newcommand{\indfun}[1]{\mathbbmss{1}\lefto\{#1\right\}}

\safemath{\matA}{\mathsf{A}}
\safemath{\matB}{\mathsf{B}}
\safemath{\matC}{\mathsf{C}}
\safemath{\matD}{\mathsf{D}}
\safemath{\matE}{\mathsf{E}}
\safemath{\matF}{\mathsf{F}}
\safemath{\matG}{\mathsf{G}}
\safemath{\matH}{\mathsf{H}}
\safemath{\matI}{\mathsf{I}}
\safemath{\matJ}{\mathsf{J}}
\safemath{\matK}{\mathsf{K}}
\safemath{\matL}{\mathsf{L}}
\safemath{\matM}{\mathsf{M}}
\safemath{\matN}{\mathsf{N}}
\safemath{\matO}{\mathsf{O}}
\safemath{\matP}{\mathsf{P}}
\safemath{\matQ}{\mathsf{Q}}
\safemath{\matR}{\mathsf{R}}
\safemath{\matS}{\mathsf{S}}
\safemath{\matT}{\mathsf{T}}
\safemath{\matU}{\mathsf{U}}
\safemath{\matV}{\mathsf{V}}
\safemath{\matW}{\mathsf{W}}
\safemath{\matX}{\mathsf{X}}
\safemath{\matY}{\mathsf{Y}}
\safemath{\matZ}{\mathsf{Z}}
\safemath{\matSigma}{\mathsf{\Sigma}}
\safemath{\matPhi}{\mathsf{\Phi}}
\safemath{\matLambda}{\mathsf{\Lambda}}
\safemath{\matDelta}{\mathsf{\Delta}}

\safemath{\randveca}{\bm{A}}
\safemath{\randvecb}{\bm{B}}
\safemath{\randvecc}{\bm{C}}
\safemath{\randvecd}{\bm{D}}
\safemath{\randvece}{\bm{E}}
\safemath{\randvecf}{\bm{F}}
\safemath{\randvecg}{\bm{G}}
\safemath{\randvech}{\bm{H}}
\safemath{\randveci}{\bm{I}}
\safemath{\randvecj}{\bm{J}}
\safemath{\randveck}{\bm{K}}
\safemath{\randvecl}{\bm{L}}
\safemath{\randvecm}{\bm{M}}
\safemath{\randvecn}{\bm{N}}
\safemath{\randveco}{\bm{O}}
\safemath{\randvecp}{\bm{P}}
\safemath{\randvecq}{\bm{Q}}
\safemath{\randvecr}{\bm{R}}
\safemath{\randvecs}{\bm{S}}
\safemath{\randvect}{\bm{T}}
\safemath{\randvecu}{\bm{U}}
\safemath{\randvecv}{\bm{V}}
\safemath{\randvecw}{\bm{W}}
\safemath{\randvecx}{\bm{X}}
\safemath{\randvecy}{\bm{Y}}
\safemath{\randvecz}{\bm{Z}}

\safemath{\randvecLambda}{\bm{\Lambda}}

\safemath{\randmatA}{\amsbb{A}}
\safemath{\randmatB}{\amsbb{B}}
\safemath{\randmatC}{\amsbb{C}}
\safemath{\randmatD}{\amsbb{D}}
\safemath{\randmatE}{\amsbb{E}}
\safemath{\randmatF}{\amsbb{F}}
\safemath{\randmatG}{\amsbb{G}}
\safemath{\randmatH}{\amsbb{H}}
\safemath{\randmatI}{\amsbb{I}}
\safemath{\randmatJ}{\amsbb{J}}
\safemath{\randmatK}{\amsbb{K}}
\safemath{\randmatL}{\amsbb{L}}
\safemath{\randmatM}{\amsbb{M}}
\safemath{\randmatN}{\amsbb{N}}
\safemath{\randmatO}{\amsbb{O}}
\safemath{\randmatP}{\amsbb{P}}
\safemath{\randmatQ}{\amsbb{Q}}
\safemath{\randmatR}{\amsbb{R}}
\safemath{\randmatS}{\amsbb{S}}
\safemath{\randmatT}{\amsbb{T}}
\safemath{\randmatU}{\amsbb{U}}
\safemath{\randmatV}{\amsbb{V}}
\safemath{\randmatW}{\amsbb{W}}
\safemath{\randmatX}{\amsbb{X}}
\safemath{\randmatY}{\amsbb{Y}}
\safemath{\randmatZ}{\amsbb{Z}}
\safemath{\randmatSigma}{\mathbb{\Sigma}}
\safemath{\randmatPhi}{\mathbb{\Phi}}

\safemath{\pdff}{f}
\safemath{\pdfp}{p}
\safemath{\pdfq}{q}

\safemath{\cdfF}{F}
\safemath{\cdfP}{P}
\safemath{\cdfQ}{Q}

\safemath{\veca}{\bm{a}}
\safemath{\vecb}{\bm{b}}
\safemath{\vecc}{\bm{c}}
\safemath{\vecd}{\bm{d}}
\safemath{\vece}{\bm{e}}
\safemath{\vecf}{\bm{f}}
\safemath{\vecg}{\bm{g}}
\safemath{\vech}{\bm{h}}
\safemath{\veci}{\bm{i}}
\safemath{\vecj}{\bm{j}}
\safemath{\veck}{\bm{k}}
\safemath{\vecl}{\bm{l}}
\safemath{\vecm}{\bm{m}}
\safemath{\vecn}{\bm{n}}
\safemath{\veco}{\bm{o}}
\safemath{\vecp}{\bm{p}}
\safemath{\vecq}{\bm{q}}
\safemath{\vecr}{\bm{r}}
\safemath{\vecs}{\bm{s}}
\safemath{\vect}{\bm{t}}
\safemath{\vecu}{\bm{u}}
\safemath{\vecv}{\bm{v}}
\safemath{\vecw}{\bm{w}}
\safemath{\vecx}{\bm{x}}
\safemath{\vecy}{\bm{y}}
\safemath{\vecz}{\bm{z}}

\safemath{\veclambda}{\bm{\lambda}}
\safemath{\vecpi}{\bm{\pi}}
\safemath{\vecsigma}{\bm\sigma}              			

\safemath{\setA}{\mathcal{A}}
\safemath{\setB}{\mathcal{B}}
\safemath{\setC}{\mathcal{C}}
\safemath{\setD}{\mathcal{D}}
\safemath{\setE}{\mathcal{E}}
\safemath{\setF}{\mathcal{F}}
\safemath{\setG}{\mathcal{G}}
\safemath{\setH}{\mathcal{H}}
\safemath{\setI}{\mathcal{I}}
\safemath{\setJ}{\mathcal{J}}
\safemath{\setK}{\mathcal{K}}
\safemath{\setL}{\mathcal{L}}
\safemath{\setM}{\mathcal{M}}
\safemath{\setN}{\mathcal{N}}
\safemath{\setO}{\mathcal{O}}
\safemath{\setP}{\mathcal{P}}
\safemath{\setQ}{\mathcal{Q}}
\safemath{\setR}{\mathcal{R}}
\safemath{\setS}{\mathcal{S}}
\safemath{\setT}{\mathcal{T}}
\safemath{\setU}{\mathcal{U}}
\safemath{\setV}{\mathcal{V}}
\safemath{\setW}{\mathcal{W}}
\safemath{\setX}{\mathcal{X}}
\safemath{\setY}{\mathcal{Y}}
\safemath{\setZ}{\mathcal{Z}}
\safemath{\emptySet}{\varnothing}

\safemath{\veczero}{\mathbf{0}} 

\safemath{\diag}{\mathrm{diag}}

\safemath{\jpg}{\mathcal{CN}}			
\safemath{\complexset}{\amsbb{C}}
\safemath{\realset}{\amsbb{R}}
\safemath{\natunum}{\mathbb{N}}
\safemath{\posrealset}{\realset_{+}}
\safemath{\integerset}{\amsbb{N}}

\newcommand{\given}{\,\vert\,}				
\safemath{\define}{\triangleq}			

\usepackage{bm}
\usepackage{upgreek}
\usepackage{amssymb}
\usepackage{amsfonts}
\usepackage{mathrsfs}
\usepackage{xspace}

\safemath{\mi}{I}
\safemath{\difent}{h}

\safemath{\constrm}{\mathrm{const}}

\safemath{\NonnegReal}{\mathbb{R}^{+}}
\safemath{\re}{\mathrm{re}}
\safemath{\Real}{\mathrm{Re}} 
\safemath{\gradient}{\nabla}
\safemath{\genericpdf}{f}

\safemath{\bl}{n} 
\safemath{\error}{\epsilon} 

\safemath{\cohtime}{n_\mathrm{c}}
\safemath{\rxant}{m_\mathrm{r}}
\safemath{\txant}{m_\mathrm{t}}

\safemath{\snr}{P}
\safemath{\const}{k}

\safemath{\spanm}{\mathrm{span}}

\safemath{\altg}{\tilde{g}}

\safemath{\altk}{\tilde{k}}
\safemath{\altconst}{\altk}
\safemath{\altvecLambda}{\widetilde{\randvecLambda}}

\safemath{\altveclambda}{\tilde{\veclambda}}
\safemath{\altvecsigma}{\tilde{\vecsigma}}              

\safemath{\altLambda}{\widetilde{\Lambda}}

\safemath{\altgamma}{\tilde{\gamma}}
\safemath{\altsigma}{\tilde{\sigma}}      				
\safemath{\altdelta}{\tilde{\delta}}
\safemath{\altrho}{\tilde{\rho}}
\safemath{\altlambda}{\tilde{\lambda}}

\safemath{\altsnr}{\altrho}						

\safemath{\altmatSigma}{\matDelta}
\safemath{\altmatPhi}{\widetilde{\matPhi}} 		
\safemath{\inpdist}{Q_{\randmatX}}

\def\eb{E_{\mathrm{b}}}

\def\cohtime{n_{\mathrm{c}}}

\def\altX{\tilde{X}}

\begin{document}
\IEEEoverridecommandlockouts

\title{A Beta-Beta Achievability   Bound with Applications}

\author{\IEEEauthorblockN{Wei Yang$^1$, Austin Collins$^2$, Giuseppe Durisi$^3$, \\Yury Polyanskiy$^2$, and H. Vincent Poor$^1$}
\thanks{
This work was supported in part by the US  National Science Foundation (NSF) under Grants  CCF-1420575 and ECCS-1343210, by the Swedish Research Council, under grant 3222452, by the Center for Science of Information (CSoI), an NSF Science and Technology Center, under grant agreement CCF-09-39370, and by the NSF CAREER award CCF-12-53205.}
\IEEEauthorblockA{
$^1$Princeton University, Princeton, NJ, 08544 USA\\
$^2$Massachusetts Institute of Technology, Cambridge, MA, 02139 USA\\
$^3$Chalmers University of Technology, 41296 Gothenburg, Sweden
\vspace{-3mm}}}

\maketitle
\begin{abstract}
A channel coding achievability bound expressed in terms of the ratio between two Neyman-Pearson~$\beta$ functions is proposed. This bound is the dual of a converse bound established earlier by Polyanskiy and Verd\'{u} (2014).  
The new bound turns out to simplify considerably  the analysis in situations where the channel output distribution is not a product distribution, for example due to a cost constraint or a structural constraint (such as orthogonality or constant composition) on the channel inputs. 
%
Connections to existing bounds in the literature are discussed. 
The bound is then used to derive 1) an achievability bound on the channel dispersion of additive non-Gaussian noise channels with random Gaussian codebooks, 2) the channel dispersion of an exponential-noise channel, 3) a second-order expansion for the minimum energy per bit of an AWGN channel, and 4) a lower bound on the maximum coding rate of a multiple-input multiple-output Rayleigh-fading channel with perfect channel state information at the receiver, which is the tightest known achievability result. 
\end{abstract}

\section{Introduction}

We consider an abstract channel that consists of an input set~$\setA$, an output set~$\setB$, and a random transformation $P_{Y|X}: \setA \to \setB$.
 An $(M,\error)$ code for the channel $(\setA, P_{Y|X},\setB)$ comprises a message set $\setM \define \{1,\ldots, M\}$, an encoder  $f: \setM \to \setA$, and a decoder $g: \setB \to \setM \cup \{e\}$ ($e$ denotes~an error event) that satisfies the \emph{average}  error probability constraint
\begin{IEEEeqnarray}{rCl}
 \frac{1}{M} \sum\limits_{j=1}^{M}  \Big(1- P_{Y\given X} \mathopen{}\big( g^{-1}(j) \given f(j) \big)\Big) \leq \error.
\label{eq:avg-prob-error-def}
\end{IEEEeqnarray}
Here, $g^{-1}(j) \define \{y\in\setY: g(y) =j\}$.
For a fixed arbitrary $\error\in(0,1)$, we are interested in finding a lower bound (i.e., an achievability bounds) on the largest number $M^*$ of codewords  for which an $(M,\error)$ code exists.

For stationary memoryless channels, Shannon's channel coding theorem establishes that  the rate of the best code   converges to the channel capacity
 \begin{IEEEeqnarray}{rCl}
 C = \max_{P_X} I(X;Y)  
 \end{IEEEeqnarray}
as the blocklength grows to infinity. Here, $I(X;Y)$ denotes the mutual information between the channel input and output.  The mutual information can be expressed through an arbitrary output distribution $Q_{Y}$ as follows~\cite[Eq.~(8.7)]{csiszar11}:
\begin{equation}
I(X;Y) =  D(P_{Y|X} \| Q_{Y}|P_X) - D(P_{Y}\| Q_{Y}).
\label{eq:golden-formula-intro}
\end{equation}
This identity---also known as the \emph{golden formula}---has found many applications in  information theory. 
For example, it allows us to prove converse bounds on channel capacity (by dropping the term $-D(P_Y\|Q_Y)$; see~\cite{lapidoth03-10a}). It is also used in the derivation of the capacity per unit cost~\cite{verdu90-09}, in the Blahut-Arimoto algorithm~\cite{arimoto1972-01a,blahut1972-07a},  in Gallager's formula for the minimax redundancy in universal source coding~\cite{gallager79-u}, and in characterizing properties of good channel codes~\cite{shamai97-03,polyanskiy14-01}.
Therefore, it is of paramount interest to find a finite-blocklength analog of~\eqref{eq:golden-formula-intro}. 

As a first step, Polyanskiy and Verd\'{u} recently proved that  every $(M,\error)$ code satisfies the following converse bound~\cite[Th.~15]{polyanskiy14-01}
\begin{IEEEeqnarray}{rCl}
M \leq \inf\limits_{0\leq \delta < 1-\error}  \inf\limits_{Q_Y} \frac{\beta_{1-\delta }(P_{Y},Q_Y)}{  \beta_{1-\error -\delta} (P_{XY} , P_{X}Q_{Y})}.
\label{eq:prop-lb-beta-p-q-intro}
\end{IEEEeqnarray}
Here, $P_{X}$ and  $P_{Y}$  denote the empirical input and output distributions induced by  the code (for the case of uniformly distributed messages). The function $\beta_\alpha(P,Q)$ in~\eqref{eq:prop-lb-beta-p-q-intro} measures the difficulty of distinguishing $P$ from~$Q$ in terms of  hypothesis testing, and is defined as\footnote{By the Neyman-Pearson lemma~\cite{neyman33a}, there exists an optimal $P_{Z|W}$ that attains the minimum in~\eqref{eq:def-beta-intro}. This test, which we shall refer to as  the Neyman-Pearson test, involves thresholding the  Radon-Nikodym derivative of $P$ with respect to $Q$. }
\begin{equation}
\label{eq:def-beta-intro}
\beta_\alpha(P,Q) \define  \min \int P_{Z\given W}(1\given w)  Q(d w)
\end{equation}
where the minimum is over all conditional probability distributions (i.e., tests) $P_{Z\given W} :\setW \to\{0,1\}$ satisfying
\begin{equation}
 \int P_{Z\given W} (1\given w) P(dw) \geq \alpha
\end{equation}
and $\setW$ denotes the support of $P$ and $Q$. The analogy between~\eqref{eq:golden-formula-intro} and~\eqref{eq:prop-lb-beta-p-q-intro} follows from Stein's lemma:
\begin{equation}
-\log \beta_{\alpha}(P^n,Q^n) = n D(P\|Q) + o(n), \quad \forall \alpha\in(0,1).
\end{equation}
%


\paragraph*{Contributions} 
In this paper, we continue the program of establishing a finite-blocklength analog of the golden formula by proving the following achievability counterpart of~\eqref{eq:prop-lb-beta-p-q-intro}. 
 
\begin{thm}[$\beta\beta$ achievability bound]
\label{thm:betabeta-bound}
For every $0<\error<1$ and every input distribution $P_{X}$, there exists an $(M, \error)$ code for the channel $(\setA, P_{Y|X}, \setB)$ satisfying
\begin{equation}
\frac{M}{2} \geq \sup\limits_{0<\tau<\error} \sup\limits_{Q_{Y}} \frac{\beta_{\tau}(P_{Y} , Q_Y )}{\beta_{1-\error+\tau}(P_{XY}, P_X Q_{Y})}
\label{eq:kappa-beta-intro-avg}
\end{equation}
where $P_Y\define P_{Y|X} \circ P_{X}$.
\end{thm}

 The proof of this bound relies on Shannon's random coding technique and on a suboptimal decoder that is based on the Neyman-Pearson test between $P_{XY}$ and $P_XQ_Y$. 
Hypothesis testing is used twice in the proof: to relate the decoding error probability to $\beta_{1-\error+\tau}(P_{XY}, P_X Q_{Y})$, and to perform a change of measure from $P_Y$ to $Q_Y$.
%
%
%

The bound~\eqref{eq:kappa-beta-intro-avg} is useful in situations where~$P_Y$ is not a product distribution (although the underlying channel law $P_{Y|X}$ is stationary and memoryless), for example  due to cost constraints, or structural constraints  on the channel input, such as orthogonality or constant composition.
In such cases, traditional achievability bounds such as Feinstein's bound~\cite{feinstein54a} and the dependence-testing (DT) bound~\cite[Th.~18]{polyanskiy10-05}, which are explicit in  $dP_{Y|X}/dP_{Y}$, become difficult to evaluate. 
In contrast, the~$\beta\beta$ bound~\eqref{eq:kappa-beta-intro-avg}  requires the evaluation of  $dP_{Y|X}/dQ_Y$, which factorizes  for product $Q_Y$. This allows for an analytical computation of~\eqref{eq:kappa-beta-intro-avg}.
Furthermore, the term $\beta_{\tau}(P_Y, Q_Y)$, which captures the cost of the change of measure from $P_Y$ to $Q_Y$, can be evaluated or bounded even when a closed-form expression for $P_Y$ is not available.
To illustrate these points, we present the following applications of Theorem~\ref{thm:betabeta-bound}:
\begin{itemize}

\item  We derive an achievability bound on the channel dispersion~\cite[Def.~1]{polyanskiy10-05} of additive non-Gaussian noise channels, for the case in which the encoder uses a power-constrained random Gaussian codebook.
We show that the power constraint introduces an additional term in the expression of the achievable dispersion, which depends on the minimum mean square error (MMSE) of estimating the channel input given the channel output. 

\item We characterize the channel dispersion of the  additive exponential noise channel introduced in~\cite{verdu96-01a}.   The channel dispersion  of a discrete couterpart of the exponential-noise channel is studied in~\cite{riedl2011-10a}.

\item We prove a second-order expansion for the minimum energy per bit of an additive white Gaussian noise (AWGN) channel at finite blocklength, hence establishing a  nonasymptotic counterpart of the ``wideband slope'' result of Verd\'{u}~\cite{verdu02-06}. Even though this result can be obtained via other techniques (such as the $\kappa\beta$ bound~\cite[Th.~25]{polyanskiy10-05}),  the proof based on~\eqref{eq:kappa-beta-intro-avg} is conceptually simpler and generalizes to other channel models.

\item We evaluate~\eqref{eq:kappa-beta-intro-avg}  for a multiple-input multiple-output (MIMO) Rayleigh-fading channel with  perfect channel state information at the receiver (CSIR).  In this case, \eqref{eq:kappa-beta-intro-avg} yields the tightest known achievability result.  
\end{itemize}


\paragraph*{Notation}
For an input distribution $P_{X}$ and a channel $P_{Y|X}$, we let $P_{Y|X}\circ P_X$ denote the distribution of $Y$ induced by $P_X$ through $P_{Y|X}$.
The distribution of a circularly symmetric complex Gaussian random vector with covariance matrix $\mathsf{A}$ is denoted by $\jpg(0,\mathsf{A})$.
With $\chi_{k}^2(\lambda)$ we denote the noncentral chi-sqared distribution with $k$ degrees of freedom and noncentrality parameter $\lambda$. 
Finally,  $\mathrm{Exp}(\mu)$ stands for the exponential distribution with mean $\mu$.

\section{Proof of Theorem~\ref{thm:betabeta-bound}}


Fix $\error \in(0,1)$, $\tau\in(0,\error)$, and let $P_X$ and $Q_{Y}$ be two arbitrary probability measures on $\setA$ and $\setB$, respectively. Furthermore, let
\begin{IEEEeqnarray}{rCl}
M= \left\lceil \frac{2\beta_{\tau}(P_{Y} , Q_Y )}{\beta_{1-\error+\tau}(P_{XY}, P_X Q_{Y})} \right\rceil.
\label{eq:def-M-ceil-ratio}
\end{IEEEeqnarray}
Finally, let $P_{Z\given X,Y}:\setA\times\setB \to\{0,1\}$ be the Neyman-Pearson test that satisfies
\begin{IEEEeqnarray}{rCl}
P_{XY}[Z(X,Y) =1] &\geq& 1-\error +\tau \label{eq:def-z-under-p-intro}\\
P_XQ_{Y}[Z(X,Y)=1] &=& \beta_{1-\error+\tau}(P_{XY}, P_X Q_{Y}).  \label{eq:def-z-under-q-intro}
\end{IEEEeqnarray}
For a given codebook $\{c_1,\ldots, c_{M}\}$ and a received signal~$y$, the decoder computes the values of $Z(c_j, y)$ and returns the smallest index $j$ for which $Z(c_j,y)=1$. If no such index is found, the decoder declares an error.
The average probability of error of the given codebook $\{c_1,\ldots, c_{M}\}$, under the assumption of uniformly distributed messages, is given by
\begin{IEEEeqnarray}{rCl}
&&P_{\mathrm{e}}(c_1,\ldots,c_M) = \prob\mathopen{}\Big[\big\{Z(c_{W}, Y) =0 \big\} \notag\\
&&\qquad \qquad  \bigcup  \big\{\exists \, m<  W \,\,\mathrm{s.t.}\,\, Z(c_m,Y)=1 \big\} \Big] \IEEEeqnarraynumspace
\label{eq:rand-coding-error-ana}
\end{IEEEeqnarray}
where $W$ is equiprobable on $\{1,\ldots,M\}$ and $Y\sim P_{Y|W}$.

Following Shannon's random coding idea, we next average~\eqref{eq:rand-coding-error-ana} over  all  codebooks $\{C_1,\ldots, C_M\}$ whose codewords are generated as pairwise independent random variables with distribution $P_X$. This yields
\begin{IEEEeqnarray}{rCl}
\IEEEeqnarraymulticol{3}{l}{\Ex{}{P_{\mathrm{e}}(C_1,\ldots,C_M) }}\notag\\
\quad &\leq& \prob\mathopen{}\big[Z(X, Y) =0 \big] + 
\prob\mathopen{}\bigg[ \! \max_{m <  W} \!Z(C_m,Y)=1   \bigg] \label{eq:double-beta-union-bound} \IEEEeqnarraynumspace\\
&\leq & \error -\tau + \prob\mathopen{}\bigg[  \max_{m < W} Z(C_m,Y)=1  \bigg] \label{eq:double-beta-ht-bound}.
\end{IEEEeqnarray}
Here,~\eqref{eq:double-beta-union-bound} follows from the union bound and~\eqref{eq:double-beta-ht-bound} follows from~\eqref{eq:def-z-under-p-intro}.

To conclude the proof of~\eqref{eq:kappa-beta-intro-avg}, it suffices to show that
\begin{IEEEeqnarray}{rCl}
\prob\mathopen{}\bigg[ \max_{m < W} Z(C_m,Y)=1  \bigg]&\leq& \tau.
\label{eq:exists-m-z-1-py}
\end{IEEEeqnarray}
Consider the randomized test $P_{\widetilde{Z}\given Y}: \setY \to \{0,1\}$:
\begin{IEEEeqnarray}{rCl}
\widetilde{Z}(y) \define \max_{m < W} Z(C_m, y).
\label{eq:def-z-widetilde-intro}
\end{IEEEeqnarray}
It follows that
\begin{IEEEeqnarray}{rCl}
\beta_{P_{Y}[\widetilde{Z} =1] } (P_{Y}, Q_{Y})&\leq & Q_{Y}[\widetilde{Z} (Y)=1]  \label{eq:def-beta-implies-betap}\\
&\leq&\frac{1}{M} \sum\limits_{j=1}^{M} (j-1) P_XQ_{Y}[Z(X,Y) =1 ] \notag\\
 \label{eq:split-qy-tildez-1} \\
&=& \frac{M-1}{2} P_X Q_{Y}[Z( X, Y)=1] \\
&=& \frac{M-1}{2} \beta_{1-\error+\tau} (P_{XY}, P_X Q_{Y}) \label{eq:split-qy-tildez-3} \IEEEeqnarraynumspace \\
&\leq & \beta_{\tau} (P_{Y}, Q_{Y}).\label{eq:split-qy-tildez-4}
\end{IEEEeqnarray}
Here,~\eqref{eq:def-beta-implies-betap} follows from~\eqref{eq:def-beta-intro};~\eqref{eq:split-qy-tildez-1} follows from~\eqref{eq:def-z-widetilde-intro} and  from the union bound;~\eqref{eq:split-qy-tildez-3} follows from~\eqref{eq:def-z-under-q-intro}; and~\eqref{eq:split-qy-tildez-4} follows from~\eqref{eq:def-M-ceil-ratio}. Since $\alpha \mapsto \beta_{\alpha}(P_{Y},Q_{Y})$ is  nondecreasing, we conclude that
\begin{IEEEeqnarray}{rCl}
P_{Y}[\widetilde{Z} =1] \leq \tau
\end{IEEEeqnarray}
which is equivalent to~\eqref{eq:exists-m-z-1-py}.


\section{Connection to Existing Bounds}

\label{sec:relation-to-existing-bounds}
We next illustrate the connection between Theorem~\ref{thm:betabeta-bound} and other achievability bounds.
\subsubsection{The $\kappa\beta$ bound} The $\kappa\beta$ bound in~\cite[Th.~25]{polyanskiy10-05} is based on Feinstein's maximal coding approach and on a suboptimal decoder similar to the one used in Theorem~\ref{thm:betabeta-bound}.
By further lower-bounding the $\kappa$ term in the $\kappa\beta$ bound using~\cite[Lemma~4]{polyanskiy11-08a}, we can relax it to the following bound: 
\begin{IEEEeqnarray}{rCl}
M \geq \sup\limits_{\tau\in(0,\error) }\sup\limits_{Q_Y} \frac{\beta_{\tau}(P_{Y|X}\circ P_X , Q_Y)}{\sup\nolimits_{x\in \setF} \beta_{1-\error + \tau}(P_{Y|X=x} , Q_{Y})} \IEEEeqnarraynumspace
\label{eq:kappa-beta-relax}
\end{IEEEeqnarray}
which holds under a \emph{maximal} error probability constraint. 
Here, $\setF\subset \setA$ denotes the permissible set of  codewords, and $P_X$ is an arbitrary distribution on $\setF$. 
The similarity between~\eqref{eq:kappa-beta-relax} and~\eqref{eq:kappa-beta-intro-avg} suggests that we can interpret the $\beta\beta$ bound as the average-error-probability counterpart of the $\kappa\beta$ bound. 
For the case in which $\beta_{\alpha}(P_{Y|X=x},Q_Y)$ does not depend on $x\in \setF$, 
by relaxing $M/2$ to $M$ in~\eqref{eq:kappa-beta-intro-avg} and by using~\cite[Lemma~29]{polyanskiy10-05} we obtain a weaker version of~\eqref{eq:kappa-beta-relax}
that  holds under the average error probability constraint. 
However, for the case in which $\beta_{\alpha}(P_{Y|X=x},Q_Y)$  does depend on $x\in \setF$,~\eqref{eq:kappa-beta-intro-avg}  can be both easier to analyze and numerically tighter than~\eqref{eq:kappa-beta-relax} (see Section~\ref{sec:mimo-bf} for an example).

\subsubsection{The dependence-testing (DT) bound} Setting $Q_Y =P_Y$ in~\eqref{eq:kappa-beta-intro-avg}, using that $\beta_{\tau}(P_Y,P_Y) = \tau$, and rearranging terms, we obtain
\begin{equation}
\error \leq   \inf_{\alpha \in(0,1)} \Big\{ 1-\alpha + \frac{M}{2} \beta_{\alpha}(P_{XY},P_XP_Y) \Big\}.
\label{eq:dt-example}
\end{equation}
Setting~$\alpha=P_{XY}[\log dP_{XY}/d(P_XP_Y) \geq \log (M/2)]$ and using the Neyman-Pearson lemma, we conclude that~\eqref{eq:dt-example} is equivalent to a slightly weakened version of the DT bound~\cite[Th.~18]{polyanskiy10-05} with $(M-1)/2$ replaced by $M/2$.
Since this weakened version of the DT bound implies Shannon's bound~\cite{shannon57} and the bound in~\cite[Th.~2]{wang2009-07a}, our bound implies these two bounds as well.

%

\ifthenelse{\boolean{conf}}
{}
{
\section{Properties of $\beta_{\alpha}(P,Q)$} 
In this section, we collect properties of $\beta_{\alpha}(P,Q)$, that are useful for evaluating~\eqref{eq:kappa-beta-intro-avg}.

\begin{lemma} 
\label{lemma:properties-of-beta}
For every $\alpha\in[0,1]$ the following hold:
\begin{enumerate}
\item \emph{Data-processing inequality~\cite{polyanskiy10-09a}:} for every stochastic kernel $P_{Y|X}: \setX\to\setY$, every $P_X$, and every~$Q_X$
\begin{equation}
\beta_{\alpha}(P_X, Q_X) \leq \beta_{\alpha}(P_{Y|X}\circ P_X, P_{Y|X} \circ Q_{X}).
\label{eq:data-processing-beta}
\end{equation}
\item \emph{Action on mixtures of distributions~\cite[Lemma~25]{polyanskiy13}:} let $P_X =\sum_j \lambda_j P_{X_j}$ be a convex combination of  $P_{X_j}$. Then for all $ Q_Y$ and $j$ we have
\begin{equation}
\beta_{\alpha}(P_{X,Y}, P_XQ_Y) \geq \lambda_j \beta_{1-(1-\alpha) \lambda^{-1}_j}(P_{X_j Y}, P_{X_j} Q_Y \!) .
\end{equation}
If the supports of $P_{X_j}$ are pairwise disjoint then
\begin{IEEEeqnarray}{rCl}
\beta_{\alpha}(P_{X,Y}, P_XQ_Y\!) &=& \!\!\! \inf\limits_{\sum_j\!\lambda_j\alpha_j  =\alpha}\!\! \!\!\! \lambda_j \beta_{\alpha_j}(P_{X_j Y} , P_{X_j} Q_Y\!).  \IEEEeqnarraynumspace
\end{IEEEeqnarray}
\item \emph{Action on product distributions:} $\forall P_1,P_2,Q_1,Q_2$
\begin{IEEEeqnarray}{rCl}
\beta_{\alpha}(P_1P_2, Q_1Q_2) 
&\geq& \beta_{\beta_{\alpha}(P_1,Q_1)} (P_2 , Q_2).
\label{eq:bound-product}
\end{IEEEeqnarray}
\item \emph{Bounds:} for every $\delta_1\in (0, 1-\alpha)$ and every $\delta_2 \in(0, \alpha)$
\begin{IEEEeqnarray}{rCl}
\frac{\beta_{\alpha +\delta_1} ( P_{XY}, P_XQ_Y)}{\beta_{\delta_1}(P_Y,Q_Y)}  &\geq&   \beta_{\alpha}(P_{XY}, P_XP_Y) \label{eq:ub-beta-q}\\
&\geq&  \frac{\beta_{\alpha - \delta_2} ( P_{XY},  P_XQ_Y)}{\gamma} \IEEEeqnarraynumspace \label{eq:lb-beta-q}
\end{IEEEeqnarray}
where $\gamma$ satisfies 
$ P_Y \mathopen{}\big[dP_Y/ dQ_Y \geq \gamma \big] \geq 1-\delta_2$.
%
\item \emph{Bounds via R\'{e}nyi divergence:} for every $\lambda>1$ 
\begin{IEEEeqnarray}{rCl}
\IEEEeqnarraymulticol{3}{l}{
\beta_{\alpha}(P,Q) }\notag\\
\,\,&& \geq \alpha^{\lambda/(\lambda-1)}\! \left( e^{(\lambda-1) D_{\lambda}(P\|Q)}  -(1-\alpha)^{\lambda}  \right)^{-1/(\lambda-1)} \IEEEeqnarraynumspace
\label{eq:bound-beta-reyni}
\end{IEEEeqnarray}
where $D_{\lambda}(\cdot\|\cdot) $ denotes the R\'{e}nyi divergence~\cite{erven2014-07a}.
\end{enumerate}
\end{lemma}
\begin{IEEEproof}
See Appendix~\ref{app:proof-lemma-properties}.
\end{IEEEproof}
}

\section{Applications}
We shall take $\setA$ and $\setB$ to be $n$-fold Cartesian products of alphabets $\setX$ and $\setY$. 
A channel is a sequence of conditional probabilities $P_{Y^n\given X^n}: \setX^n \to \setY^n$. 
We shall refer to an $(M,\error)$ code for the channel $\{\setX^n, P_{Y^n\given X^n}, \setY^n\}$ as an $(n,M,\error)$ code.
Furthermore, the maximum coding rate  $R^*(n,\error)$  is defined as\footnote{Unless otherwise indicated, the $\log$ and $\exp$ functions in this paper are taken with respect to an arbitrary fixed basis.}
\begin{equation}
R^*(n,\error) \define \sup \lefto\{\frac{\log M}{n} : \exists (n,M,\error) \text{ code}\right\}.
\end{equation}
\ifthenelse{\boolean{conf}}
{Due to space limitations,  we have omitted the proofs of all theorems in the section. They can be found in~\cite{yang16-inprep}.}
{}

\subsection{Additive non-Gaussian noise channels}

We consider the additive-noise channel
\begin{equation}
Y_i = X_i + Z_i, \quad i=1, \ldots, n
\label{eq:channel-io-nong}
\end{equation}
where $\{Z_i\}$ are independent and identically distributed (i.i.d.)   $P_Z$-distributed (not necessarily Gaussian) and $X_i,Y_i,Z_i \in \realset$.
%
%
Each codeword $x^n$ must satisfy the constraint
\begin{equation}
\|x^n\|^2 =  \sum\limits_{i=1}^{n} x_i^2 \leq n \snr .
\label{eq:power-constraint-nongaussian}
\end{equation}
Let $Q_{X^n}= \mathcal{N}(\mathbf{0}, \snr \matI_n)$, and let $P_{X^n}$ denote the conditional distribution of $X^n \sim Q_{X^n}$ conditioned on  
\begin{equation} 
X^n \in \setA_n \define \Big\{x^n \in \realset^n: n\snr - \log n \leq \|x^n\|^2 \leq n\snr\Big\} .
\label{eq:cost-constraint22}
\end{equation}
In other words, $P_{X^n}$ is a truncated Gaussian distribution that is supported on the spherical shell $\setA_n$. 
We shall consider an ensemble of codes in which the codewords are generated independently from the distribution $P_{X^n}$.  
This ensemble of codes is used by Gallager to derive the random coding error exponent for channels with cost constraint~\cite[p.~326]{gallager68}.  
In the following theorem we present a second-order asymptotic expansion on  the maximum rate achievable with this  code over the channel~\eqref{eq:channel-io-nong}.

\begin{thm}
\label{thm:additive-nonGaussian}
Let $Q_X = \mathcal{N}(0,P)$, let $Q_{Y} = P_{Y|X} \circ Q_X$, and let
\begin{equation}
i(x;y) \define \frac{dP_{Y|X}}{dQ_{Y}} (x;y) 
\end{equation}
be the information density of the joint distribution $Q_XP_{Y|X}$.
Furthermore, let 
\begin{equation}
I(\snr) \define  \Ex{Q_XP_{Y|X}}{  i(X;Y)  }.
\end{equation}
Assume that the noise $Z$ satisfies the following conditions:
\begin{enumerate}
\item $P_Z$ is  absolutely continuous with respect to the Lebesgue measure on $\realset$;
\item $\Ex{Q_{X}P_Z}{ | i(X;X+Z) -I(P) |^3} <\infty$; and
\item $\Ex{}{|Z|^6} <\infty $.
\end{enumerate}
Then, for every $0<\error<1$, we have
\begin{equation}
R^*(n,\error) \geq  I(\snr) -\sqrt{\frac{V(\snr)}{n}}Q^{-1}(\error) + \bigO\lefto(\frac{\log n}{n}\right).
\label{eq:rate-thm-exp}
\end{equation}
Here, $Q^{-1}(\cdot)$ denotes the inverse of the Gaussian $Q$-function, 
\begin{equation} 
V(\snr) \define \mathrm{Var}[ i(X;Y) |X] + \mathrm{Var}\lefto[D(P_{Y|X=\bar{X}} \| Q_Y) + c \bar{X}^2\right]  \IEEEeqnarraynumspace
\label{eq:def-var-nongaussian}
\end{equation} 
where the second variance is taken with respect to $\bar{X}\sim Q_X$,
\begin{equation}
c\define \frac{\log e}{2P^2}\Big(\mathsf{mmse}(X|Y) -P\Big)
\label{eq:thm-nongaussian-defc}
\end{equation}
and 
\begin{equation}
\mathsf{mmse}(X|Y)\define \Ex{}{(X - \Ex{}{X|Y})^2}.
\label{eq:thm-nongaussian-defmmse}
\end{equation}
In both~\eqref{eq:def-var-nongaussian} and~\eqref{eq:thm-nongaussian-defmmse}, the pair  $(X,Y)$ is distributed according to $Q_XP_{Y|X}$.
\end{thm}

\ifthenelse{\boolean{conf}}
{}
{\begin{IEEEproof}
See Appendix~\ref{app:proof-nongaussian}. 
\end{IEEEproof}}
\begin{rem}
For $P_Z = \mathcal{N}(0,1)$,~\eqref{eq:def-var-nongaussian} recovers the  dispersion $V(P  ) = \frac{P(2+P)}{2(1+P)^2} \log^2 e$ of the AWGN channel~\cite[Th.~54]{polyanskiy10-05}. 
\end{rem}


 \subsection{The exponential-noise channel}
We next consider the exponential-noise case, i.e., $P_Z =\mathrm{Exp}(1)$.  As in~\cite{verdu96-01a}, we assume that each codeword $x^n \in \realset^n$ must satisfy
\begin{equation}
 x_i \geq 0  \quad \text{ and } \quad \sum\limits_{i=1}^{n} x_i \leq n\sigma.
\label{eq:exp-cost-constraint}
\end{equation}
The practical relevance of such a channel is discussed in~\cite{verdu96-01a, anantharam1996-01a}.
The capacity of the exponential-noise channel with constraint~\eqref{eq:exp-cost-constraint} is given by~\cite[Th.~3]{verdu96-01a}
\begin{equation}
C(\sigma) = \log(1+\sigma)
\end{equation}
and is achieved by the input distribution
$P^*_X$, according to which $X$ takes the value zero with probability $1/(1+\sigma)$ and follows an $\mathrm{Exp}(1+\sigma)$ distribution conditioned on it being positive.
Furthermore, the capacity-achieving output distribution is $\mathrm{Exp}(1+\sigma)$.

\begin{thm}
\label{thm:exp-noise}
For the additive-exponential noise channel  subject to the  constraint~\eqref{eq:exp-cost-constraint} and for $0<\error<1$, 
\begin{equation}
R^*(n,\error) = \log(1+\sigma) -\sqrt{\frac{V(\sigma)}{n}}Q^{-1}(\error) + \bigO\lefto(\frac{\log n}{n}\right)
\label{eq:rate-thm-exp}
\end{equation}
where
\begin{equation}
V(\sigma) = \frac{\sigma^2}{(1+\sigma)^2} \log^2 e.  
\end{equation} 
\end{thm}
\ifthenelse{\boolean{conf}}
{}
{
\begin{IEEEproof}
See Appendix~\ref{app:proof-thm-expnoise}. 
\end{IEEEproof}
}

%
%
%
%
%

\subsection{Minimum energy per bit over AWGN channels}
For a complex-valued AWGN channel, we set $\setA  = \complexset^{n}$, $\setB  = \complexset^n$, and $P_{Y^n \given X^n =x^n } = \jpg(x^n, \matI_n)$.  
We assume that every codeword  $x^n$ satisfies the equal power constraint
\begin{equation}
\|x^n\|^2 = n\snr.
\label{eq:power-constraint}
\end{equation}
Let $R_{\mathrm{e}}^*(n,\error,\snr)$ denote the maximum coding rate $R^*(n,\error)$ under the constraint~\eqref{eq:power-constraint}.
%
In Theorem~\ref{thm:app-betabeta-awgn} below, we provide expressions for the $\beta$ functions in~\eqref{eq:prop-lb-beta-p-q-intro} and~\eqref{eq:kappa-beta-intro-avg} for the AWGN case.
%
%
%
%
%
%
%
%

%
%

\begin{thm}
\label{thm:app-betabeta-awgn}
Consider the complex-valued AWGN channel $P_{Y^n|X^n}$. 
Let $S_n \sim \chi^2_{2n}(2nP)$ and $L_n \sim \chi^2_{2n}(0)$.  Let $Q_{Y^n} = \jpg(\mathbf{0}, \matI_n)$. 
Furthermore,  let $\setS_n \define \{ x^n\in \complexset^n: \|x^n\|^2 =nP \} $. 
Then, for every distribution $P_{X^n}$ supported on~$\setS_n$
\begin{equation}
\beta_{\alpha}(P_{X^nY^n},P_{X^n} Q_{Y^n}) =  Q\lefto(\sqrt{2n \snr} + Q^{-1}(\alpha) \right)
\label{eq:beta-alpha-pxy-q}
\end{equation}
and 
\begin{equation}
\beta_{a} (P_{Y^n|X^n} \circ P_{X^n}, Q_{Y^n} ) \leq  \prob[L_n \geq \gamma] 
\label{eq:beta-py-qy-thm}
\end{equation}
where $\gamma$ satisfies 
\begin{equation}
\prob[S_n \geq \gamma ] = a.
\label{eq:def-gamma}
\end{equation}
Furthermore,~\eqref{eq:beta-py-qy-thm} holds with equality if $P_{X^n}$ is the uniform distribution over $\setS_n$.
\end{thm}

\ifthenelse{\boolean{conf}}{}
{\begin{IEEEproof}
See Appendix~\ref{app:proof-thm-beta-awgn-nonasy}.
\end{IEEEproof}}


By evaluating~\eqref{eq:beta-alpha-pxy-q} and~\eqref{eq:beta-py-qy-thm} in the asymptotic regime $\snr \to 0$ and $n\snr^2 \to \infty$ as $n\to\infty$,\footnote{As we shall see, this regime is of interest for the characterization of the minimum energy per bit.} and by substituting the resulting expressions in Theorem~\ref{thm:betabeta-bound} and in~\eqref{eq:prop-lb-beta-p-q-intro}, we obtain the following result. 

\begin{thm}
\label{thm:energy-per-bit}
For an AWGN channel with SNR $\snr_n$ satisfying $\snr_n \to 0$ and $n\snr_n^2 \to \infty$ as $n\to\infty$, the maximum coding rate  $R^*_{\mathrm{e}}(n,\error,\snr_n)$ behaves as
\begin{IEEEeqnarray}{rCl}
\frac{R^*_{\mathrm{e}}(n,\epsilon,\snr_n)}{\log e} &=&  \snr_n - \sqrt{ \frac{2 \snr_n}{n}}Q^{-1}(\error) - \frac{1}{2}\snr_n^2 \notag  \\
&&+ \littleo\mathopen{}\bigg(\sqrt{\frac{\snr_n}{n}}\bigg) + \littleo\lefto(\snr_n^2\right), \quad n\to\infty.
\label{eq:asy-expansion-r-awgn-eq}
\end{IEEEeqnarray}
\end{thm}
\ifthenelse{\boolean{conf}}{}{
\begin{IEEEproof}
See Appendix~\ref{app:proof-thm-beta-awgn-asy}.
\end{IEEEproof}}

\begin{figure}
\begin{centering}
\includegraphics[scale=0.61]{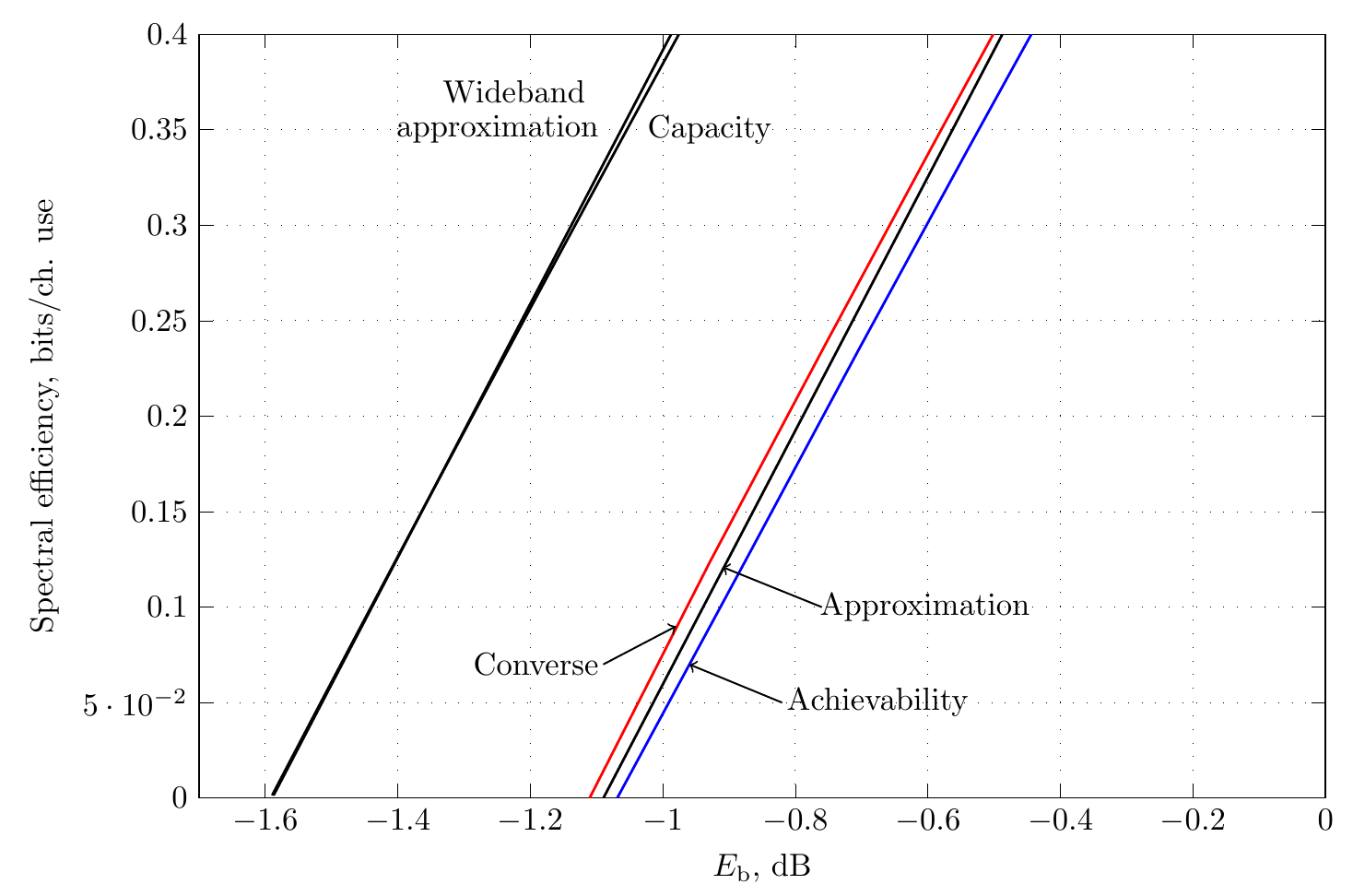}
\caption{Minimum energy per bit versus spectral efficiency of the AWGN channel; here, $k=2000$ bits, and $\error =10^{-3}$.\label{fig:minimum-energy-per-bit}}
\end{centering}
\end{figure}

We now relate~\eqref{eq:asy-expansion-r-awgn-eq} to the minimum energy per bit $\eb^*(k,\error, R)$ to transmit $k$ information bits at rate $R$ and error probability~$\error$. 
Specifically, Theorem~\ref{thm:energy-per-bit} implies that
\begin{IEEEeqnarray}{rCl}
\IEEEeqnarraymulticol{3}{l}{10\log_{10} \eb^*(k, \error, R ) }\notag\\
 &=&10\log_{10} \frac{\snr_n}{R^*_{\mathrm{e}}(n,\epsilon,\snr_n)}  \\
&=& 10 \log_{10}\mathopen{}\bigg(\log_e 2 +\sqrt{\frac{2\log_e 2}{k}} Q^{-1}(\error) + \frac{\log_e^2 2}{2}R\bigg) \notag \\
&&+\, o(R) + o(1/\sqrt{k})\\
&=& 10\log_{10}\! \eb^*(k, \error, 0)  + \frac{10\log_{10} \!2}{2} R + \littleo(R) + \littleo\lefto(\!\frac{1}{\sqrt{k}}\!\right). \IEEEeqnarraynumspace
\label{eq:wide-band-slope-sergio-awgn}
\end{IEEEeqnarray} 
The last step follows from~\cite[Th.~3]{polyanskiy11-08b}. 
Note that~\eqref{eq:wide-band-slope-sergio-awgn} is the finite-blocklength counterpart of Verd\'{u}'s \emph{wideband-slope} result~\cite[Eqs.~(172)]{verdu02-06}. 
In Fig.~\ref{fig:minimum-energy-per-bit}, we present a comparison between the approximation~\eqref{eq:wide-band-slope-sergio-awgn} (with the $\littleo(\cdot)$ term omitted), the converse bound~\cite[Th.~28]{polyanskiy10-05}, and the achievability bound~\eqref{eq:kappa-beta-intro-avg}. In both cases $Q_{Y}$ is chosen to be the capacity-achieving output distribution. For the parameters consider in Fig.~\ref{fig:minimum-energy-per-bit},  the approximation~\eqref{eq:wide-band-slope-sergio-awgn} is accurate.

\subsection{MIMO block-fading channel with perfect CSIR}
\label{sec:mimo-bf}
Consider an $\txant\times \rxant$ Rayleigh MIMO block-fading channel  that stays constant for~$\cohtime$ channel uses. The input-output relation within the $k$th coherence interval is given by 
\begin{equation}
\label{eq:channel-io-mimo}
\randmatY_k =\randmatX_k \randmatH_k + \randmatW_k.
\end{equation}
Here, $\randmatX_k\in \complexset^{\cohtime \times\txant}$ and $\randmatY_k\in\complexset^{\cohtime \times \rxant}$ are the transmitted and received matrices, respectively; the entries of the fading matrix $\randmatH_k\in\complexset^{\txant\times\rxant}$ and the noise $\randmatW_k \in\complexset^{\cohtime\times\rxant}$ are i.i.d. $\jpg(0,1)$. We assume that $\{\randmatH_k\}$ and $\{\randmatW_k\}$ take on independent realizations over successive coherence intervals. 
The channel matrices  $\{\randmatH_k\}$ are assumed to be known to the receiver but not to the transmitter.
We shall also assume that each codeword spans $l\in\amsbb{N}$ coherence intervals, i.e., the blocklength of the code is $n=l \cohtime$.
Finally, each codeword~$\matX^l$ is constrained to satisfy
\begin{equation}
\label{eq:power-mimo-constraint}
\fnorm{\matX^l} \leq \sqrt{nP}.
\end{equation}

\begin{figure}
\begin{centering}
\includegraphics[scale=0.66]{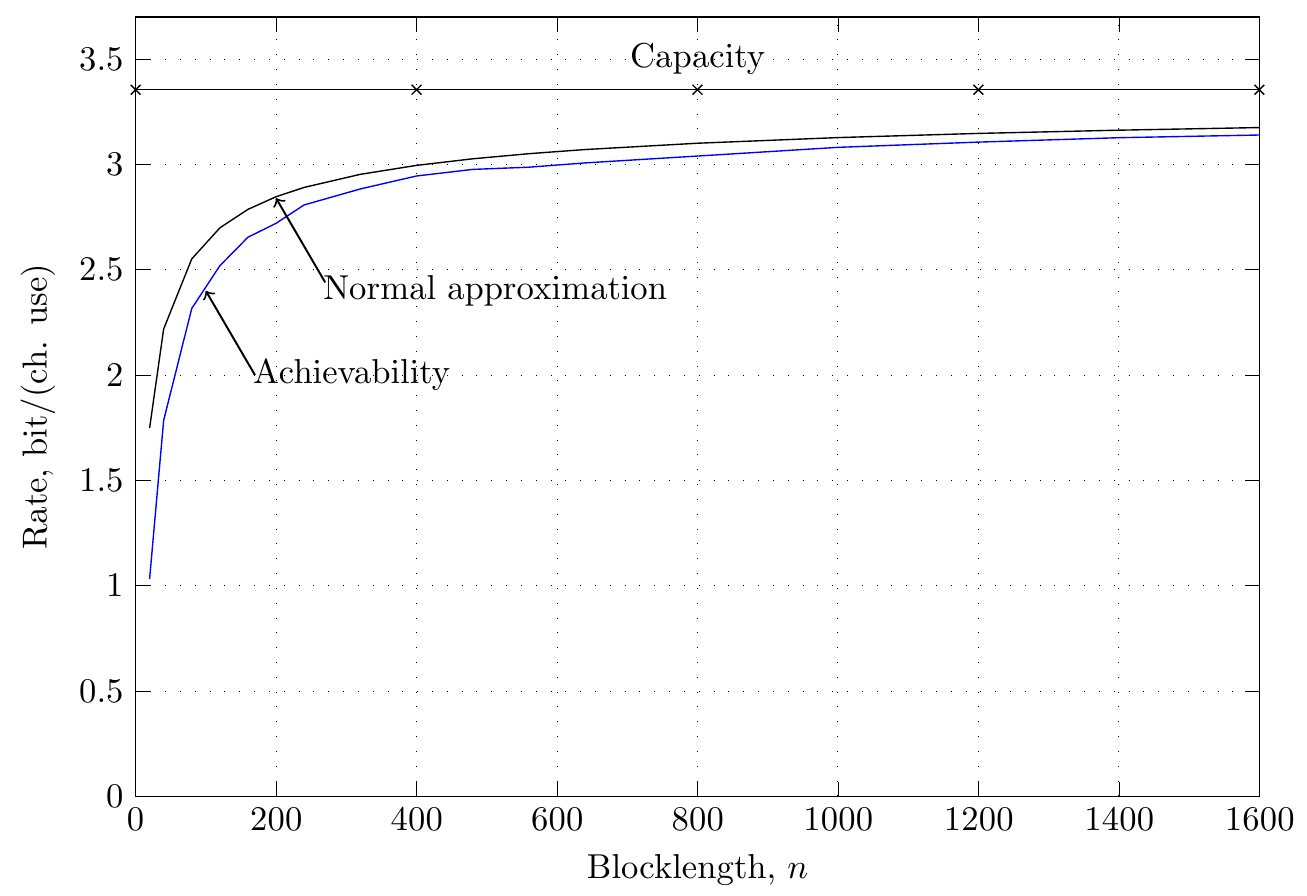}
\caption{Bounds on rate for a $4\times 4$ MIMO Rayleigh block-fading channel; here SNR=$0$ dB, $\error=0.001$, and $\cohtime =4$.\label{fig:mimo}}
\end{centering}
\end{figure}

To obtain an achievability bound on $R^*(n,\error)$, we apply Theorem~\ref{thm:betabeta-bound}
with $P_{\randmatX^l}$ chosen as the uniform distribution on $\setS_n' \define \{\matX^l: \fnorm{\matX^l}^2 = n \snr \}$ and  $Q_{\randmatY^l \randmatH^l}$ chosen as the capacity-achieving output distribution. 
With these choices, we have
\begin{equation}
R^*(n,\error) \geq \frac{1}{n}\log \frac{\beta_{\tau}(P_{\randmatH^l\randmatY^l}, Q_{\randmatH^l\randmatY^l})}{\beta_{1-\error+\tau}(P_{\randmatX^l\randmatH^l\randmatY^l},P_{\randmatX^l}Q_{\randmatH^l\randmatY^l})}.
\label{eq:mimo-betabeta-app}
\end{equation}
The denominator  $\beta_{1-\error+\tau}(P_{\randmatX^l\randmatH^l\randmatY^l},P_{\randmatX^l}Q_{\randmatH^l\randmatY^l})$ in~\eqref{eq:mimo-betabeta-app} can be computed via standard Monte Carlo techniques. 
However, computing~$\beta_{\tau}(P_{\randmatH^l\randmatY^l}, Q_{\randmatH^l\randmatY^l})$ in the numerator is more involved, since there is no closed-form expression for $P_{\randmatH^l \randmatY^l}$.
\ifthenelse{\boolean{conf}}{
 To circumvent this, we further lower-bound $\beta_{\tau}(P_{\randmatH^l\randmatY^l}, Q_{\randmatH^l\randmatY^l})$ using the data-processing inequality~\cite{polyanskiy10-09a} for $\beta_{\alpha}$ as follows. }{
 To circumvent this, we further lower-bound $\beta_{\tau}(P_{\randmatH^l\randmatY^l}, Q_{\randmatH^l\randmatY^l})$ using the data-processing inequality for $\beta_{\alpha}$ as follows.}
Let~$\widetilde{\randmatX}^l$ be a sequence of $\cohtime\times \txant$ complex matrices with i.i.d. $\jpg(0,P/\txant)$ entries. Then, $P_{\randmatX^l}$ can be obtained via  
$\widetilde{\randmatX}^l$ through $\randmatX^l = \sqrt{nP}  \widetilde{\randmatX}^l/ \big\|\widetilde{\randmatX}^l\big\|_{\mathsf{F}}$.
Furthermore, $Q_{\randmatH^l\randmatY^l} =  P_{\randmatY^l \randmatH^l \given \randmatX^l}\circ P_{\widetilde{\randmatX}^l}$. 
Let $P^{(s)}_{\randmatY^l \randmatH^l\given \randmatX^l } \define P_{\randmatH^l} P^{(s)}_{\randmatY^l \given \randmatH^l,\randmatX^l}$, where $P^{(s)}_{\randmatY^l \given \randmatH^l,\randmatX^l}$ denotes  the channel law defined by
\begin{equation}
\label{eq:equivalent-channel}
\randmatY_k =  \randmatX_k\randmatH_k  \frac{\sqrt{nP}}{\fnorm{\randmatX^l}} +\randmatW_k,\qquad k=1,\ldots, l.
 \end{equation}
We have that $P_{\randmatY^l\randmatH^l} = P_{\randmatY^l \randmatH^l \given \randmatX^l} \circ P_{\randmatX^l} = P^{(s)}_{\randmatY^l \randmatH^l\given f\randmatX^l}\circ P_{\widetilde{\randmatX}^l}$.
Now, by the data-processing inequality, 
\begin{equation}
\label{eq:data-processing-beta-app}
\beta_{\tau}(P_{\randmatH^l\randmatY^l}, Q_{\randmatH^l\randmatY^l}) \geq \beta_{\tau}( P_{\widetilde{\randmatX}^l} P^{(s)}_{\randmatY^l \randmatH^l\given \randmatX^l} ,  P_{\widetilde{\randmatX}^l}  P_{\randmatY^l \randmatH^l\given \randmatX^l} ).
\end{equation}
Since the Radon-Nikodym derivative $\frac{d(P_{\widetilde{\randmatX}^l} P^{(s)}_{\randmatY^l \randmatH^l\given \randmatX^l} ) }{d (P_{\widetilde{\randmatX}^l}  P_{\randmatY^l \randmatH^l\given \randmatX^l} )} $ can be computed in closed form, the right-hand side of~\eqref{eq:data-processing-beta-app} can be computed via Monte Carlo techniques. 
The resulting bound is compared with the normal approximation of $R^*(n,\error)$ in Fig.~\ref{fig:mimo}. 
In contrast,  the $\kappa\beta$ bound~\cite[Th.~25]{polyanskiy10-05} with $\setF = \setS_n'$ is much more difficult to compute due to the maximization over codewords $\matX^l\in \setS_n'$. Furthermore, for  blocklength values of practical interest,  we expect that
\begin{IEEEeqnarray}{rCl}
\max\limits_{\matX^l \in \setS_n'} \! \beta_{\alpha}( P_{\randmatH^l\randmatY^l\given \randmatX^l =\matX^l}, Q_{\randmatH^l\randmatY^l}) \gg \beta_{\alpha}(P_{\randmatX^l\randmatH^l\randmatY^l},P_{\randmatX^l}Q_{\randmatH^l\randmatY^l}) \IEEEeqnarraynumspace 
\end{IEEEeqnarray}
which means that the resulting bound is much looser than~\eqref{eq:mimo-betabeta-app}.


\ifthenelse{\boolean{conf}}
{}
{
\begin{appendix}

\subsection{Proof of Lemma~\ref{lemma:properties-of-beta}}
\label{app:proof-lemma-properties}
To prove~\eqref{eq:bound-product}, we assume without loss of generality that $P_1P_2$ is supported on $\setX\times\setY$ and so is $Q_1Q_2$. 
Let $Z: \setX\times\setY \to \{0,1\}$ denote the Neyman-Pearson test between $P_1P_2$ and $Q_1Q_2$, i.e.,
\begin{IEEEeqnarray}{rCl}
P_1P_2[Z(X,Y) =1] &=&\alpha\\
 Q_1Q_2[Z(X,Y)=1] &=& \beta_\alpha(P_1P_2 , Q_1Q_2). 
\label{eq:def-nptest-proof}
\end{IEEEeqnarray}
Interpreting $Z(\cdot,Y)$ as a randomized  test between $P_1$ and $Q_1$ and $Z(X, \cdot)$  between $P_2$ and $Q_2$, we obtain 
\begin{IEEEeqnarray}{rCl}
Q_1Q_2[Z(X,Y)=1] &\geq& \beta_{Q_1P_2[Z(X,Y)=1]}(P_2, Q_2) \\
&\geq & \beta_{\beta_{\alpha}(P_1,Q_1) }(P_2, Q_2).
\label{eq:lower-bound-q1q2}
\end{IEEEeqnarray}
The last step follows by the monotonicity of $\alpha \mapsto \beta_\alpha(P_2,Q_2)$.
Combining~\eqref{eq:def-nptest-proof} and~\eqref{eq:lower-bound-q1q2}, we conclude~\eqref{eq:bound-product}.

We next prove the first inequality in~\eqref{eq:ub-beta-q}. Let $Z_1:\setX\times\setY \to \{0,1\}$ be the Neyman-Pearson test that achieves $\beta_{\alpha-\delta_1}(P_{XY}, P_XQ_Y)$,  and let $Z_2:\setY \to \{0,1\}$ be the Neyman-Pearson test that achieves $\beta_{\delta_1}(P_{Y},Q_Y)$. 
Consider the following test $Z:\setX\times\setY \to \{0,1\}$:
\begin{equation}
Z(x,y) \define Z_1(x,y) \cdot (1- Z_2(y)).
\end{equation}
We have:
\begin{IEEEeqnarray}{rCl}
P_{XY}[Z=1] &=& P_{XY}[Z_1=1 , Z_2 =0 ]\\
&\geq& P_{XY}[Z_2=0] -P_{XY}[Z_1 =0] =\alpha. \IEEEeqnarraynumspace
\label{eq:lower-bound-pxy-z1}
\end{IEEEeqnarray}
Let 
\begin{equation}
\tilde{\gamma} \define \sup\{\gamma: P_Y[dP_Y/dQ_Y \geq \gamma] \geq \delta_1\}.
\label{eq:def-gamma-tilde}
\end{equation}
By the Neyman-Pearson lemma, we have $\frac{dP_Y}{dQ_Y}(y) \geq \tilde{\gamma}$ for every~$y$ such that $Z_2(y)=1$.
Hence,
\begin{IEEEeqnarray}{rCl}
\beta_{\alpha}(P_{XY},P_XP_Y) &\leq&  \beta_{P_{XY}[Z=1]}(P_{XY},P_XP_Y) \label{eq:bound-beta-pxy-1}\\
&\leq& P_XP_Y[Z=1]\\
&\leq & \tilde{\gamma}P_XQ_Y[Z_1 =1]  \\
&=& \tilde{\gamma} \beta_{\alpha + \delta_1} (P_{XY},P_XQ_Y)  \label{eq:bound-beta-pxy-2} \\
&\leq& \frac{\beta_{\alpha + \delta_1}(P_{XY},P_XQ_Y) }{\beta_{\delta_1}(P_Y,Q_Y)}.\label{eq:bound-beta-pxy-3}
\end{IEEEeqnarray}
Here,~\eqref{eq:bound-beta-pxy-1} follows from~\eqref{eq:lower-bound-pxy-z1} and because $\alpha\mapsto \beta_{\alpha}(\cdot,\cdot)$ is monotonically nondecreasing; ~\eqref{eq:bound-beta-pxy-2} follows from the definition of the test $Z_1$; finally,~\eqref{eq:bound-beta-pxy-3} follows from~\eqref{eq:def-gamma-tilde} and~\cite[Eq.~(103)]{polyanskiy10-05}. 
The proof of the second inequality in~\eqref{eq:lb-beta-q} follows from similar steps and is omitted. 

Finally, to prove~\eqref{eq:bound-beta-reyni}, we use the data-processing inequality for R\'{e}yni divergence~\cite{erven2014-07a} and view the Neyman-Pearson test between $P$ and $Q$ as a map from  the support of $P$ and $Q$ to $\{0,1\}$. This yields 
\begin{IEEEeqnarray}{rCl}
D_{\lambda}(P\|Q) &\geq& d_\lambda(\alpha \| \beta_{\alpha}) \\
&=&\frac{1}{\lambda-1}\log\lefto(\alpha^{\lambda}\beta_{\alpha}^{1-\lambda} + (1-\alpha)^\lambda (1-\beta_{\alpha})^{1-\lambda}\right)\IEEEeqnarraynumspace
\end{IEEEeqnarray}
where $d_{\lambda}(\cdot\|\cdot)$ denote the binary R\'{e}nyi divergence~\cite[Eq.~(23)]{polyanskiy10-09a}, and where $\beta_{\alpha}\define \beta_{\alpha}(P,Q)$. 
By further lower-bounding $(1-\beta_{\alpha})^{1-\lambda}$ by $1$ (recall that $\lambda >1$), and solving the resulting inequality for $\beta_{\alpha}$, we obtain~\eqref{eq:bound-beta-reyni}.

\subsection{Proof of Theorem~\ref{thm:additive-nonGaussian}}
\label{app:proof-nongaussian}

Let $P_{Y^n } \define P_{Y^n\given X^n} \circ P_{X^n} $ and let $Q_{Y^n} \define P_{Y^n\given X^n} \circ Q_{X^n}$. %
We will apply Theorem~\ref{thm:betabeta-bound} to the channel~\eqref{eq:channel-io-nong} with $\tau = 1/\sqrt{n}$ and with $P_{X^n}$ and $Q_{Y^n}$ chosen as above. 
For the term $\beta_{\tau}(P_{Y^n},Q_{Y^n})$, we have 
\begin{IEEEeqnarray}{rCl}
\IEEEeqnarraymulticol{3}{l}{
 \log \beta_{\tau}(P_{Y^n},Q_{Y^n})} \notag\\ 
\,\, &\geq& \log \beta_{\tau}(P_{X^n} , Q_{X^n})\label{eq:lower-bound-beta-1}\\
&= &  \log Q_{X^n}\lefto[ n\snr - \log n \leq  \sum\limits_{i=1}^{n} X_i^2 \leq n\snr \right] + \log \tau  \label{eq:lower-bound-beta-2} \IEEEeqnarraynumspace\\
&=& \log \lefto(Q\lefto( \frac{ - \log n}{\sqrt{2n P^2}}\right)  - Q(0) - \bigO\lefto(\frac{1}{\sqrt{n}} \right)\!\right) +\log \tau \label{eq:lower-bound-beta-3} \IEEEeqnarraynumspace\\
&=&  \bigO\mathopen{}\big(\log n\big) \label{eq:lower-bound-beta-4}.
\end{IEEEeqnarray}
Here, \!\eqref{eq:lower-bound-beta-1} follows from the data-processing inequality \eqref{eq:data-processing-beta};~\eqref{eq:lower-bound-beta-2} follows by using the Neyman-Pearson lemma and by observing that $dP_{X^n}/dQ_{X_n} $ is a binary random variable;~\eqref{eq:lower-bound-beta-3} follows from the Berry-Esseen theorem~\cite[Sec. XVI.5]{feller70a}.

Let $p_Z$ and $q_Y$ be the probability density function (pdf)  corresponding to the  distributions $P_Z$ and $Q_Y$, respectively. 
Since $Q_X$ is a Gaussian distribution,  it follows that $q_{Y}(\cdot)$ is smooth. 
To evaluate $\beta_{1-\error+\tau} (P_{X^n Y^n},P_{X^n} Q_{Y^n})$, we shall prove that,
under $P_{X^nY^n}$, the random variable 
\begin{equation} 
\log\frac{dP_{Y^n| X^n} }{dQ_{Y^n}} (X^n, Y^n) \sim \sum_{i=1}^{n} \Big( \log p_Z(Z_i) -  \log q_Y(X_i + Z_i) \Big) 
\label{eq:info-density-dist-nong}
\end{equation}
is asymptotically normal as $n\to\infty$. The main difficulty in establishing this result is that, under $P_{X^n}$,  the $\{X_i\}$ are not independent, which prevents us from using the Berry-Esseen theorem.
To circumvent this difficulty, we introduce a joint probability distribution $P_{X^n,\altX^n}$  with marginals $X^n\sim P_{X^n}$ and $\altX^n\sim Q_{X^n}$, under which $\Ex{}{\|X^n - \altX^n\|^2}$ is small, and we approximate $ \log q_Y(X_i + Z_i)$ by $\log q_Y(\altX_i +Z_i)$.
Specifically, let $P_{X^n,\altX^n}$ be the joint distribution for which  $X^n/\|X^n\| =\altX^n/\|\altX_n\| $ and for which $\|X^n\|$ is independent of $\|\altX^n\|$.
%
%
%
Writing $\Delta_i \define \altX_i -X_i$ and $f(y) \define \log q_{Y}(y)$, we next relate $f(X_i + Z_i)$ to $f(\altX_i +Z_i)$ as follows:\footnote{In this appendix, statements involving``$=$'', ``$\geq$'', and ``$\leq$'' hold with probability one, whenever random variables are involved.}
\begin{IEEEeqnarray}{rCl}
\IEEEeqnarraymulticol{3}{l}{
  f(\altX_i +Z_i) - f(X_i + Z_i)  } \notag\\
\quad &=& \int\nolimits_{0}^{1} \Delta_i f'(Z_i + X_i + \Delta_i t) dt\\
&=& \Delta_i f'(\altX_i +Z_i)  - \Delta_i^2  \int\nolimits_{0}^{1} t  f''(X_i +Z_i + t \Delta_i) dt  \IEEEeqnarraynumspace 
\label{eq:relate-fx-fxt}
\end{IEEEeqnarray} 
where in the last step we used  integration by parts. 
The second-order derivative $f''(y)$ can be computed  as follows (see~\cite[Eq.~(131)]{wu2012-03a})
\begin{IEEEeqnarray}{rCl}
f''(y) = \frac{d^2 \log q_{Y} (y)}{dy^2} &=& -\frac{\log e }{P} + \frac{\log e}{P^2} \mathrm{Var}[X| Y=y]  \label{eq:bound-mmse-x-1}\IEEEeqnarraynumspace\\
&\leq&  -\frac{\log e }{P} +  \frac{2\log e }{P^2}  ( y^2 + \mathrm{Var}[Z]) \IEEEeqnarraynumspace \label{eq:bound-mmse-x}
\end{IEEEeqnarray}
where the conditional variance on the right-hand side (RHS) of~\eqref{eq:bound-mmse-x-1} is computed with respect to $(X,Y)\sim Q_XP_{Y|X}$. Here,~\eqref{eq:bound-mmse-x}
 follows from steps similar to the ones leading to~\cite[Eq.~(169)]{wu2012-03a}. 
 Substituting~\eqref{eq:bound-mmse-x} into~\eqref{eq:relate-fx-fxt} and upper-bounding~$t$ by $1$, we obtain  
\begin{IEEEeqnarray}{rCl}
  f(X_i + Z_i)  
  &\leq&    f(\altX_i +Z_i)  - \Delta_i f'(\altX_i +Z_i) \notag\\
  &&+  \underbrace{\frac{\log e}{P}  \Delta_i^2 \left( c_0+ \frac{2}{P}\! \int \nolimits_{0}^{1} \!(X_i +Z_i + t\Delta_i)^2  dt \right)  }_{ \define T_{1,i}}\notag\\
\label{eq:bound-f-x-xt}
\end{IEEEeqnarray}
where  $c_0 \define |2 \mathrm{Var}[Z] /P  -1| $. 
Furthermore,  
\begin{IEEEeqnarray}{rCl}
\Delta_i &= &\altX_i -X_i \\
&=& \frac{\altX_i}{\|\altX^n\|}\lefto(\|\altX^n\| - \sqrt{nP} \right) +   \frac{\altX_i}{\|\altX^n\|}\lefto( \sqrt{nP} - \|X^n\| \right)  \IEEEeqnarraynumspace
\label{eq:rewirte-Delta-i}
\end{IEEEeqnarray}
where in the last step we used that $\altX^n / \|\altX^n\| = X^n /\| X^n\|$, and that both $\altX^n / \|\altX^n\| $ and $X^n/\|X^n\|$ are independent of $\|X\|^n$.
Letting  
\begin{IEEEeqnarray}{rCl}
A_{1,i} &\define& \log p_Z(Z_i) -  f(\altX_i + Z_i)= i(\altX_i, \altX_i+Z_i)\IEEEeqnarraynumspace \\
T_{2,i} &\define&  f'(\altX_i +Z_i) \frac{\altX_i}{\|\altX^n\|}\lefto( \sqrt{nP} - \|X^n\| \right) 
\end{IEEEeqnarray}
and substituting~\eqref{eq:bound-f-x-xt} and~\eqref{eq:rewirte-Delta-i} into~\eqref{eq:info-density-dist-nong}, we obtain
\begin{IEEEeqnarray}{rCl}
\IEEEeqnarraymulticol{3}{l}{
 \sum_{i=1}^{n} \Big( \log p_Z(Z_i) -  \log q_Y(X_i + Z_i) \Big)  }\notag\\
&\geq &  \sum_{i=1}^{n} \! \bigg( \!A_{1,i}   +  f'(\altX_i +Z_i)\altX_i \bigg(\!1- \frac{\sqrt{nP}}{\|\altX^n\|}\bigg) - T_{1,i} + T_{2,i} \!  \bigg). \notag\\
\label{eq:final-bound-on-ixy}
\end{IEEEeqnarray}
Using~\eqref{eq:final-bound-on-ixy} together with the inequality
\begin{equation}
\prob[A+B\leq a+b] \leq \prob[A\leq a] + \prob[B\leq b]
\label{eq:inequality-a-b}
\end{equation}
we  conclude that for every  $\gamma$, $\gamma_1$, and $\gamma_2$,
\begin{IEEEeqnarray}{rCl}
\IEEEeqnarraymulticol{3}{l}{
P_{X^nY^n}\lefto[ \log\frac{dP_{Y^n| X^n} }{dQ_{Y^n}} (X^n, Y^n) \leq \gamma - \gamma_1 - \gamma_2\right] } \notag\\
&\leq & \prob\mathopen{}\bigg[ \sum_{i=1}^{n} \! A_{1,i}   +  f'(\altX_i +Z_i)\altX_i \bigg(1- \frac{\sqrt{nP}}{\|\altX^n\|}\bigg)   \leq \gamma  \bigg] \notag\\
&& +\,\prob\lefto[ \sum_{i=1}^{n} T_{1,i}   \geq \gamma_1 \right] +  \prob\lefto[- \sum_{i=1}^{n} T_{2,i} \geq  \gamma_2 \right].
\label{eq:bound-prob-ixy-gamma}
\end{IEEEeqnarray}

We next uper-bound the three probability terms on the RHS of~\eqref{eq:bound-prob-ixy-gamma} separately. To bound the first term, we shall use the central-limit theorem for functions~\cite[Prop.~1]{molavianJazi15-12a} (see also~\cite[Prop.~1]{iri2015-06a}) and follow similar steps as in~\cite[Sec.~IV.D]{molavianJazi15-12a}. Let
\begin{IEEEeqnarray}{rCl}
A_{2,i} &\define& f'(\altX_i + Z_i) \altX_i \label{eq:def-a2i} \label{eq:def-a2i}\\
 A_{3,i} &\define&  \altX_i^2 - \snr.
\end{IEEEeqnarray}
It follows that  $\{[A_{1,i}, A_{2,i} ,A_{3,i}]\}$ are i.i.d. random vectors. Note also that 
\begin{IEEEeqnarray}{rCl}
\Ex{}{A_{1,i}} &=& I(P),\quad \Ex{}{A_{3,i}} = 0\label{eq:compute-A-1-3}
\end{IEEEeqnarray}
and that
\begin{IEEEeqnarray}{rCl}
\Ex{}{A_{2,i}} 
 &=& \Ex{}{f'(\altX_1 +Z_1) \altX_1}\\
&=&\int x p_X(x) p_Z(z) \frac{\log e}{P}
 \Ex{ }{-X | Y=x+z} dxdz \IEEEeqnarraynumspace \label{eq:compute-d-logy1}\\
 &=&  - \frac{\log e}{P}\!\!\int x \tilde{x} p_X(x) p_Z(z) p_{X|Y}(\tilde{x}|x+z) d\tilde{x} dxdz \label{eq:compute-d-logy2}\IEEEeqnarraynumspace\\
 &=&  -\frac{\log e}{P} \!\!\int\! x\tilde{x} p_X(x) p_{Y|X}(x|y) p_{X|Y}(\tilde{x}|y)  d\tilde{x} dxdy\label{eq:compute-d-logy3} \IEEEeqnarraynumspace\\
 &=& -\frac{\log e}{P} \Ex{Y}{ \amsbb{E}^2\lefto[ X|Y \right] } \label{eq:compute-d-logy4}\\
 &=& \frac{\log e}{P} \Big(\mathsf{mmse}(X|Y) - P\Big) \label{eq:compute-d-logy5}.
 \end{IEEEeqnarray}
The  conditional expectations in~\eqref{eq:compute-d-logy1} and~\eqref{eq:compute-d-logy4} are computed with respect to $(X,Y)\sim Q_XP_{Y|X}$, the functions $p_X$, $p_{Y|X}$, and $p_{X|Y}$ are the (conditional) pdfs corresponding to $Q_X$,  $P_{Y|X}$, and $P_{X|Y}$, respectively, and $\mathsf{mmse}(X|Y)$ in~\eqref{eq:compute-d-logy5} is  defined in~\eqref{eq:thm-nongaussian-defmmse}.
Here,~\eqref{eq:compute-d-logy1}  follows from~\cite[Eq.~(15)]{polyanskiy2015-04a}; in~\eqref{eq:compute-d-logy3} we have used the change of variables $y=x+z$ and that $p_{Y|X}(y|x) = p_Z(y-x)$; and finally~\eqref{eq:compute-d-logy5} follows because $\Ex{Q_{X}}{X^2} = P$.
Furthermore,  we have 
\begin{IEEEeqnarray}{rCl}
\Ex{}{|A_{2,i} |^3} &\leq& \Ex{}{|\altX_i|^3 (c_1 |\altX_{1,i} +Z_i| +c_2 )^3} 
\label{eq:bound-A2-3}  \\
&\leq& {c_3} \Ex{}{|\altX_i|^6} + c_4 \Ex{}{|\altX_i|^3}\Ex{}{|Z_i|^3} <\infty  \label{eq:bound-A2-3-2} \IEEEeqnarraynumspace
\end{IEEEeqnarray}
where $c_1,c_2,c_3,c_4$ are finite constants. Here,~\eqref{eq:bound-A2-3} follows from~\eqref{eq:def-a2i} and~\cite[Prop.~2]{polyanskiy2015-04a}, and~\eqref{eq:bound-A2-3-2} follows because   $\Ex{}{|Z_i|^3}\leq \sqrt{\Ex{}{|Z_i|^6}} < \infty$.
Since $\Ex{}{|A_{1,i} -I(P)|^3} <\infty$ by assumption and since $\Ex{}{|A_{3,i}|^3}<\infty$, we have verified that the vector $[A_{1,i}, A_{2,i} , A_{3,i}]$ has finite third  central moment. 
Let now
\begin{equation}
g(a_1,a_2,a_3) \define a_1 + a_2 \left(1- \sqrt{ \frac{P}{P+a_3}}\right).
\end{equation}
We have
\begin{IEEEeqnarray}{rCl}
\IEEEeqnarraymulticol{3}{l}{
g\lefto( \frac{1}{n} \sum\limits_{i=1}^n A_{1,i}  ,  \frac{1}{n} \sum\limits_{i=1}^n A_{2,i}   , \frac{1}{n} \sum\limits_{i=1}^n A_{3,i}   \right)  }\notag\\ 
\,\, &=& \frac{1}{n} \sum_{i=1}^{n}\!  \bigg( A_{1,i}   +  f'(\altX_i +Z_i)\altX_i \bigg(1- \frac{\sqrt{nP}}{\|\altX^n\|}\bigg)  \bigg) .\IEEEeqnarraynumspace
\end{IEEEeqnarray}
Let $\vecj $ denote the Jacobian of $g$ at $(I(P),\Ex{}{A_{2,i}},0)$, and let $\matV$ denote the covariance matrix of $[A_{1,1}, A_{2,1} , A_{3,1}]$. It follows that
\begin{IEEEeqnarray}{rCl}
\vecj =[1 , 0, c] 
\end{IEEEeqnarray}
where $c$ is defined in~\eqref{eq:thm-nongaussian-defc}. Furthermore, 
\begin{IEEEeqnarray}{rCl}
\vecj \matV \tp{\vecj} &=& 
\mathrm{Var}[A_{1,1}] + 2c \cdot \mathrm{Cov}[A_{1,1} , \altX_1^2] + c^2\mathrm{Var}[\altX_1^2] \IEEEeqnarraynumspace\\
&=& \mathrm{Var}[A_{1,1} + c \altX_1^2 ]\\
&=& \mathrm{Var}[  i(\altX_1 ; \altX_1 + Z_1)| \altX_1] \notag\\
&&+\, \mathrm{Var}_{\altX_1}\lefto[ \Ex{Z_1}{  i(\altX_1 ;\altX_1 + Z_1) } + c \altX_1^2\right] \IEEEeqnarraynumspace\\
& =& V(P)
\end{IEEEeqnarray}
where $V(P)$ is defined in~\eqref{eq:def-var-nongaussian}. 
We now invoke~\cite[Prop.~1]{iri2015-06a} and obtain
\begin{IEEEeqnarray}{rCl}
\IEEEeqnarraymulticol{3}{l}{
 \prob\mathopen{}\bigg[ \sum_{i=1}^{n} \! A_{1,i}   +  f'(\altX_i +Z_i)\altX_i \bigg(1- \frac{\sqrt{nP}}{\|\altX^n\|}\bigg)   \leq \gamma  \bigg]  }\notag\\
\quad &\leq& Q\lefto( \frac{ nI(P) -\gamma }{\sqrt{nV(P)}} \right) + \bigO\lefto(\frac{1}{\sqrt{n}}\right). 
\label{eq:berry-esseen-func-1}
\end{IEEEeqnarray}

We next upper-bound the second term on the RHS of~\eqref{eq:bound-prob-ixy-gamma}. 
Note first that, by~\eqref{eq:cost-constraint22}, 
\begin{IEEEeqnarray}{rCl}
 0&\leq& \sqrt{n\snr} - \|X^n\|\\
  & \leq& \sqrt{n\snr} -\sqrt{n\snr -\log n} \leq \tilde{c}_1 n^{-1/2}\log n
 \label{eq:bound-Gamma}
 \end{IEEEeqnarray}
 for some $\tilde{c}_1> 0$ and sufficiently large $n$. 
Evaluating the integral on the RHS of~\eqref{eq:bound-f-x-xt}, using~\eqref{eq:rewirte-Delta-i} and~\eqref{eq:bound-Gamma}, and using that $(a+b)^2\leq 2(a^2 + b^2)$, we obtain
\begin{IEEEeqnarray}{rCl}
\frac{P}{\log e}T_{1,i} &=& \Delta_i^2 \Big( c_0+ \frac{2}{P}\big(\altX_i + Z_i - \Delta_i/2\big)^2 + \frac{\Delta_i^2}{6P} \Big) \\
&\leq &\underbrace{2\left(1 -\frac{\sqrt{nP} }{\|\altX^n\|}\right)^2 }_{\define T_{3}} \underbrace{ {\altX_i^2}\left(c_0+ \frac{4}{P}(\altX_i+Z_i)^2\right)}_{\define T_{4,i}} \notag\\
&&+ \underbrace{\frac{7}{6P} \Delta_i^4 + \frac{2\tilde{c}_1^2 \altX_i^2}{\|\altX^n\|^2} \frac{\log^2 n}{n}\!\Big( c_0+ \frac{4}{P}\big(\altX_i + Z_i\big)^2 \!\Big)}_{\define T_{5,i}} .\notag\\
\end{IEEEeqnarray}
Set $\gamma_1 = (\tilde{c}_2\tilde{c}_3 + 1 ) (\log n)\cdot (\log e) / P$, where $\tilde{c}_2$ and $\tilde{c}_3$ are sufficiently large constants. We have
\begin{IEEEeqnarray}{rCl}
\prob\lefto[\sum\limits_{i=1}^{n} T_{1,i} \geq\gamma_1 \right] &\leq& 
\prob\lefto[ T_3 \geq \frac{\tilde{c}_2 \log n}{n}\right] + \prob\lefto[\sum\limits_{i=1}^{n} T_{4,i} \geq \tilde{c}_3 n\right] \notag\\
&& + \, \prob\lefto[\sum\limits_{i=1}^{n}T_{5,i}\geq \log n\right] \label{eq:bound-sum-t1}
\end{IEEEeqnarray}
which follows by a repeated use of~\eqref{eq:inequality-a-b}.
The first term on the RHS of~\eqref{eq:bound-sum-t1} is upper-bounded by
\begin{IEEEeqnarray}{rCl}
\IEEEeqnarraymulticol{3}{l}{
 \prob\lefto[ T_3 \geq \frac{\tilde{c}_2 \log n}{n}\right] }\notag\\
 \quad &\leq& \prob\lefto[ \|\altX^n\| \geq \frac{n\sqrt{P}}{\sqrt{n}-\sqrt{(\tilde{c}_2/2) \log n }})\right]\IEEEeqnarraynumspace \notag \\
&&+\, \prob\lefto[ \|\altX^n\| \leq \frac{n\sqrt{P}}{\sqrt{n}+\sqrt{(\tilde{c}_2/2) \log n }})\right].\label{eq:bound-t3-0} \IEEEeqnarraynumspace
\end{IEEEeqnarray}
The first term on the RHS of~\eqref{eq:bound-t3-0}  can be further upper-bounded as
\begin{IEEEeqnarray}{rCl}
\IEEEeqnarraymulticol{3}{l}{
\prob\lefto[ \|\altX^n\| \geq \frac{n\sqrt{P}}{\sqrt{n}-\sqrt{(\tilde{c}_2/2) \log n }})\right]
}\notag\\
\quad &\leq& \prob\lefto[\|\altX^n\|^2 \geq nP(1+ \sqrt{\tilde{c}_2 (\log n) /(2n)})^2 \right] \label{eq:bound-t3-1} \\
&\leq& Q\lefto( \sqrt{\tilde{c}_2 \log n} + \frac{\tilde{c}_2 \log n}{2\sqrt{2n}}\right) + \bigO\lefto(\frac{1}{\sqrt{n}}\right) \label{eq:bound-t3-2}\\
&=& \bigO\lefto(\frac{1}{\sqrt{n}}\right) \label{eq:bound-t3-3}.
\end{IEEEeqnarray}
Here, in~\eqref{eq:bound-t3-1}  we used that $1/(1-x) \geq (1+x)$ for every $x\in(0,1)$;~\eqref{eq:bound-t3-2} follows from the Berry-Esseen theorem;~\eqref{eq:bound-t3-3} follows by using $Q(x) \leq e^{-x^2/2 }$ and by taking $\tilde{c}_2$ sufficiently large.

We can upper-bound the second term on the RHS of~\eqref{eq:bound-t3-0} using similar methods and obtain
\begin{IEEEeqnarray}{rCl}
 \prob\lefto[ T_3 \geq \frac{\tilde{c}_2 \log n}{n}\right] \leq \bigO\lefto(\frac{1}{\sqrt{n}}\right) .\label{eq:bound-t3}
\end{IEEEeqnarray}
Furthermore, we can upper-bound the second term on the RHS of~\eqref{eq:bound-sum-t1} using again the  Berry-Esseen theorem and   the assumption $\Ex{}{|Z_i|^6}<\infty$. Doing so we obtain that, for sufficiently large $\tilde{c}_3$,
\begin{IEEEeqnarray}{rCl}
\prob\lefto[\sum\limits_{i=1}^{n} T_{4,i} \geq \tilde{c}_3 n\right] \leq \bigO\lefto(\frac{1}{\sqrt{n}}\right) .\label{eq:bound-t4}
\end{IEEEeqnarray}
Finally, to bound the last term on the RHS of~\eqref{eq:bound-sum-t1}, we observe that  
\begin{IEEEeqnarray}{rCl}
\Ex{}{\sum\limits_{i=1}^n T_{5,i}} = \bigO\lefto(\frac{\log^2 n}{n}\right)
\label{eq:bound-t5-mean}
\end{IEEEeqnarray}
which follows from algebraic manipulations. Since $T_{5,i}\geq 0$, by Markov's inequality, this implies that
\begin{IEEEeqnarray}{rCl}
\prob\lefto[\sum\limits_{i=1}^n T_{5,i}  \geq \log n \right] &\leq& \bigO\lefto( \frac{\log n}{n}\right).\label{eq:bound-t5}
\end{IEEEeqnarray}
Substituting~\eqref{eq:bound-t3},~\eqref{eq:bound-t4}, and~\eqref{eq:bound-t5} into~\eqref{eq:bound-sum-t1}, we conclude that 
\begin{IEEEeqnarray}{rCl}
\prob\lefto[\sum\limits_{i=1}^{n} T_{1,i} \geq \gamma_1\right] \leq  \bigO\lefto(\frac{1}{\sqrt{n}}\right).
\label{eq:bound-second-term-prob}
\end{IEEEeqnarray}

To bound the third term on the RHS of~\eqref{eq:bound-prob-ixy-gamma}, we notice that, by the analysis in~\eqref{eq:def-a2i}--\eqref{eq:berry-esseen-func-1}, the random variable 
\begin{equation}
\|\altX^n\|^{-1}\sum\limits_{i=1}^{n} f'(\altX_i +Z_i)\altX_i  - \sqrt{n} \frac{\Ex{}{A_{2,1}}  }{\sqrt{P}}
\end{equation}
converges in distribution to $\sqrt{V'}Z'$, where $Z'\sim \mathcal{N}(0,1)$, for some finite $V'>0$. 
Furthermore, the speed of convergence is $O(1/\sqrt{n})$. 
Letting $\Gamma \define \sqrt{nP} -\|X^n\| $  and using that $\Gamma$ is independent of $\altX^n$, we obtain 
\begin{IEEEeqnarray}{rCl}
 \prob\mathopen{}\bigg[ - \sum_{i=1}^{n} T_{2,i} \geq  \gamma_2 \bigg]  &= &\Ex{}{ Q\lefto(\!\frac{ \gamma_2 /\Gamma + \sqrt{n/P}\Ex{}{A_{2,1}}}{\sqrt{V'}}\!\right)\!} \notag\\ &&+ \, \bigO\lefto(\frac{1}{\sqrt{n}}\right).
 \label{eq:bound-third-term-prob}
\end{IEEEeqnarray}
 Setting $\gamma_2 = \tilde{c}_4  \log n$ and using~\eqref{eq:bound-Gamma}, we get 
\begin{IEEEeqnarray}{rCl}
\IEEEeqnarraymulticol{3}{l}{
Q\lefto(\frac{ \gamma_2 /\Gamma + \sqrt{n}\Ex{}{A_{2,1}}}{\sqrt{V'}}\right) }\notag\\
 \quad &\leq& Q\lefto(\sqrt{n} \frac{ \tilde{c}_4/\tilde{c}_1 + \Ex{}{A_{2,1}}/\sqrt{P} }{\sqrt{V'}}\right) \IEEEeqnarraynumspace \\
 &=& \bigO(e^{-n})
 \label{eq:bound-q-third}
\end{IEEEeqnarray}
provided that $\tilde{c}_4 >  - \tilde{c}_1 E[A_{2,1}]/\sqrt{P}$ (recall that $\Ex{}{A_{2,1}}\leq 0$, see~\eqref{eq:compute-d-logy5}).

Finally, substituting~\eqref{eq:bound-q-third} into~\eqref{eq:bound-third-term-prob}, and then~\eqref{eq:berry-esseen-func-1},~\eqref{eq:bound-second-term-prob}, and~\eqref{eq:bound-third-term-prob} into~\eqref{eq:bound-prob-ixy-gamma}, we obtain
\begin{IEEEeqnarray}{rCl}
\IEEEeqnarraymulticol{3}{l}{
P_{X^nY^n}\lefto[ \log\!\frac{dP_{Y^n| X^n} }{dQ_{Y^n}} (X^n, Y^n) \leq \gamma -  \gamma_1-\gamma_2\right] } \notag\\
\quad &\leq & Q\lefto(\frac{n I(P) -\gamma }{\sqrt{nV(P)}}\right) +\bigO\mathopen{}\left(\frac{1}{\sqrt{n}}\right).
\label{eq:bound-prob-ixy-gamma-final}
\end{IEEEeqnarray}
Setting $\gamma \define n I(P) - \sqrt{n V(P)} Q^{-1}(\error - \tilde{c}_5/\sqrt{n})$ and recalling that $\tau = 1/\sqrt{n}$,  we conclude that the left-hand side of~\eqref{eq:bound-prob-ixy-gamma-final} is upper-bounded by $\error  - \tau$ for sufficiently large  $\tilde{c}_5$. By the standard upper bound~\cite[Eq.~(103)]{polyanskiy10-05} on $\beta_{\alpha}(P,Q)$, this implies   
\begin{IEEEeqnarray}{rCl}
\IEEEeqnarraymulticol{3}{l}{
- \log \beta_{1-\error+\tau} (P_{X^n Y^n},P_{X^n} Q_{Y^n}) }\notag\\
\quad &\geq& \gamma -  \gamma_1-\gamma_2 \\
 &=& n I(P) - \sqrt{n V(P)} Q^{-1}(\error) + \bigO(\log n).
\label{eq:upper-bound-beta-deno-nonga}
\end{IEEEeqnarray}
We conclude the proof of~\eqref{eq:rate-thm-exp} by combining~\eqref{eq:lower-bound-beta-4} and~\eqref{eq:upper-bound-beta-deno-nonga} with~\eqref{eq:kappa-beta-intro-avg}.

\subsection{Proof of Theorem~\ref{thm:exp-noise}}
\label{app:proof-thm-expnoise}
Let $Q_{X^n}= (P_{X}^*)^n$, and as in Theorem~\ref{thm:additive-nonGaussian} we shall choose $P_{X^n}$ as the conditional distribution of $X^n\sim Q_{X^n}$ given that
\begin{equation}
n\sigma - \log n \leq  \sum_{i=1}^{n} X_i \leq n\sigma .
\label{eq:cost-constraint22}
\end{equation}
By construction, $X^n\sim P_{X^n}$ satisfies the constraint~\eqref{eq:exp-cost-constraint}.
Finally, let $P_{Y^n } \define P_{Y^n\given X^n} \circ P_{X^n} $ and let $Q_{Y^n} \define P_{Y^n\given X^n} \circ Q_{X^n}$. %
We will apply Theorem~\ref{thm:betabeta-bound} to the exponential-noise channel with $\tau = 1/\sqrt{n}$ and with $P_{X^n}$ and $Q_{Y^n}$ chosen as above. 
As in~\eqref{eq:lower-bound-beta-1}--\eqref{eq:lower-bound-beta-4}, we bound $\beta_{\tau}(P_{Y^n},Q_{Y^n})$ using the data-processing inequality and the Berry-Esseen Theorem. This yields:
\begin{IEEEeqnarray}{rCl}
 \log \beta_{\tau}(P_{Y^n},Q_{Y^n})  
&\geq &  \bigO\mathopen{}\big(\log n\big) \label{eq:lower-bound-beta-n}.
\end{IEEEeqnarray}
For the exponential-noise channel,    $\beta_{1-\error+\tau} (P_{X^n Y^n},P_{X^n} Q_{Y^n})$ is easy to compute. 
Indeed,   under $P_{X^nY^n}$, the random variable $\log\frac{dP_{Y^n| X^n} }{dQ_{Y^n}} (X^n, Y^n)$ has the same distribution as 
\begin{IEEEeqnarray}{rCl}
n\log (1+ \sigma)   + \frac{\log e}{1+\sigma}\sum\limits_{i=1}^{n} X_i - \frac{\sigma\log e}{1+\sigma}\sum\limits_{i=1}^{n} Z_i . \IEEEeqnarraynumspace
\label{eq:info-density-dist-exp}
\end{IEEEeqnarray}
This random variable depends on the codeword $X^n$ only through $\sum\nolimits_{i=1}^{n} X^n$.
Furthermore, given $\sum\nolimits_{i=1}^{n} X^n$, this random variable is the sum of $n$ i.i.d. random variables. 
Using  the Berry-Esseen theorem and~\eqref{eq:cost-constraint22} to evaluate~\eqref{eq:info-density-dist-exp}, we conclude that 
\begin{IEEEeqnarray}{rCl}
\IEEEeqnarraymulticol{3}{l}{
\log \beta_{1-\error+\tau} (P_{X^n Y^n},P_{X^n} Q_{Y^n}) }\notag\\
\quad  &=& n\log(1+P) -\sqrt{nV(\sigma)}Q^{-1}(\error) + \bigO(\log n). \IEEEeqnarraynumspace
\label{eq:CLT-beta-deno-exp-n}
\end{IEEEeqnarray}
Substituting~\eqref{eq:CLT-beta-deno-exp-n} and~\eqref{eq:lower-bound-beta-n} into~\eqref{eq:kappa-beta-intro-avg}, we establish that~\eqref{eq:rate-thm-exp} is achievable. The converse  follows from the meta-converse theorem~\cite[Th.~27]{polyanskiy10-05} and from~\eqref{eq:CLT-beta-deno-exp-n}.

\subsection{Proof of Theorem~\ref{thm:app-betabeta-awgn}}
\label{app:proof-thm-beta-awgn-nonasy}

Due to the spherical symmetry of $\setS_n$ and $Q_{Y^n}$, we have that $\beta_{\alpha}(P_{Y^n|X^n =x^n}, Q_{Y^n} )$ is independent of $x^n\in\setS_n$. 
Let $\bar{x}^n \define [\sqrt{nP},0,\ldots,0]$. By~\cite[Lemma~29]{polyanskiy10-05}, for every $P_{X^n}$ supported on $\setS_n$,
\begin{IEEEeqnarray}{rCl}
\beta_{\alpha}(P_{X^nY^n},P_{X^n} Q_{Y^n}) &=& \beta_{\alpha}(P_{Y^n\given X^n =\bar{x}^n }, Q_{Y^n}) \\
&=& \beta_{\alpha}(\jpg(\sqrt{nP}, 1), \jpg(0,1)). \IEEEeqnarraynumspace
\label{eq:beta-pq-gaussian-01}
\end{IEEEeqnarray} 
We obtain~\eqref{eq:beta-alpha-pxy-q} by applying the Neyman-Pearson lemma to the RHS of~\eqref{eq:beta-pq-gaussian-01}. 

To prove~\eqref{eq:beta-py-qy-thm}, it suffices to  observe that $Z(y^n) = \indfun{2\|y^n\|^2 \geq \gamma }$ defines a test between $P_{Y^n}$ and $Q_{Y^n}$. Furthermore, under $P_{Y^n}$, the random variable $2\|Y^n\|^2$ has the same distribution as $S_n$ regardless of $P_{X^n}$, and under $Q_{Y^n}$, it has the same distribution as $L_n$.

It remains to show that~\eqref{eq:beta-py-qy-thm} holds with equality if $P_{X^n}$ is the uniform distribution over $\setS_n$. 
In this case, both $P_{Y^n}$ and $Q_{Y^n}$ are isotropic, and, hence, $2\|Y\|^n$ is a sufficient statistics for testing between $P_{Y^n}$ and $Q_{Y^n}$. Let $p_0$ and $p_1$ denote the pdfs of $\chi_{2n}^{2}(2n\snr)$ and  $\chi_{2n}^{2}(0)$, respectively.
 Following simple algebraic manipulations, it can be shown that $t \mapsto \frac{p_0(t)}{p_1(t)}$ is monotonically nondecreasing on $(0,\infty)$. Hence, the test $Z(y^n) = \indfun{2\|y\|^n \geq \gamma }$ coincides with the Nayman-Pearson test between $\chi_{2n}^{2}(2nP)$ and $\chi_{2n}^{2}(0)$. The desired result follows then from the Neyman-Pearson lemma.

\subsection{Proof of Theorem~\ref{thm:energy-per-bit}}
\label{app:proof-thm-beta-awgn-asy}
%
Choosing $P_{X^n}$ to be the uniform distribution over $\setS_n$,  substituting~\eqref{eq:beta-alpha-pxy-q} and~\eqref{eq:beta-py-qy-thm} into~\eqref{eq:kappa-beta-intro-avg},  and setting $a = \tau_n \define \max\{(1+3\sqrt{2})n^{-1/2} , e^{-nP_n^2/2}\}$, we obtain
\begin{IEEEeqnarray}{rCl}
R_{\mathrm{e}}^*(n,\error,\snr_n) &\geq& \frac{1}{n} \log \frac{\prob[L_n\geq \gamma] }{  Q\lefto(\sqrt{2n \snr_n} + Q^{-1}(1-\epsilon + \tau_n) \right)  } \IEEEeqnarraynumspace
\label{eq:beta-beta-awgn-app1}
\end{IEEEeqnarray}
where $L_n$ and $\gamma$ are defined in Theorem~\ref{thm:app-betabeta-awgn}.
Using the  expansion of the $Q$-function given in~\cite[Eq.~(3.53)]{verdu98}, we obtain that as $nP_n \to \infty$
\begin{IEEEeqnarray}{rCl}
\IEEEeqnarraymulticol{3}{l}{
 -\log Q\lefto(\sqrt{2n \snr_n} + Q^{-1}(1-\error+\tau_n) \right) }\notag \\
 \quad  &=& n \snr_n\log e - \sqrt{2n\snr_n}Q^{-1}(\error )\log e + \littleo\lefto(\sqrt{n \snr_n}\right). \IEEEeqnarraynumspace
 \label{eq:denominator-beta-beta-awgn}
\end{IEEEeqnarray}
Next, we compute the numerator of~\eqref{eq:beta-beta-awgn-app1}. 
Since $S_n$ in~\eqref{eq:def-gamma} has the same distribution as $\sum\nolimits_{i=1}^{2n} (Z_i + \sqrt{P_n})^2$ with $Z_i \sim \mathcal{N}(0,1)$, we can estimate $\prob[S_n \geq \gamma]$ using the Berry-Esseen theorem~as
\begin{IEEEeqnarray}{rCl}
\left| \prob[S_n \geq \gamma] - Q\lefto( \frac{\gamma - 2n(1+ P_n) }{ \sqrt{4n(1+ 2P_n)}} \! \right) \right| &\leq& \frac{2^{3/2}(1+3P_n) }{(1+2P_n)^{3/2}\sqrt{n}} \IEEEeqnarraynumspace \label{eq:app-berry-esseen-awgn}\\
& \leq& {k_1}/{\sqrt{n}}
\end{IEEEeqnarray}
where $k_1 \define 2\sqrt{2}$. Using~\eqref{eq:def-gamma} and recalling $a = \tau_n$, we obtain
\begin{equation}
\gamma \leq 2n(1+ P_n) + \sqrt{4n(1+ 2P_n)} Q^{-1}(\tau_n - k_1/\sqrt{n})\define \tilde{\gamma}.
\label{eq:def-gamma_tilde}
\end{equation}
Furthermore, we have 
\begin{IEEEeqnarray}{rCl}
\log \prob[L_n \geq \gamma] &\geq& \log \prob[L_n \geq \tilde{\gamma}]\\
&\geq &  -\frac{1}{2}\Big\{ (\tilde{\gamma} - 2n)\log e\notag\\
&&  + (2n-2) \log \frac{ \tilde{\gamma}}{2n} +\log(2n)\Big\}  -\log 2 \label{eq:bound-cdf-chi-squared} \IEEEeqnarraynumspace\\
 &=&  - \frac{1}{2} n P_n^2 \log e + \littleo(nP_n^2), \quad nP_n^2 \to \infty. \IEEEeqnarraynumspace 
 \label{eq:bound-cdf-chi-squared2}
\end{IEEEeqnarray}
Here,~\eqref{eq:bound-cdf-chi-squared} follows from~\cite[Lemma~5]{inglot06-09a} and because $L_n\sim \chi^2_{2n}(0)$, and~\eqref{eq:bound-cdf-chi-squared2} follows from~\eqref{eq:def-gamma_tilde} and from algebraic manipulations. Substituting~\eqref{eq:denominator-beta-beta-awgn} and~\eqref{eq:bound-cdf-chi-squared2} into~\eqref{eq:beta-beta-awgn-app1}, we conclude that~\eqref{eq:asy-expansion-r-awgn-eq} is achievable.

To prove the converse, we substitute~\eqref{eq:beta-alpha-pxy-q} and~\eqref{eq:beta-py-qy-thm} into~\eqref{eq:prop-lb-beta-p-q-intro}, and set $a = \alpha_n = 1-\tau_n$. 
Doing so we obtain 
\begin{equation}
R_{\mathrm{e}}^*(n,\error,\snr_n) \leq  \frac{1}{n} \log \frac{\prob[L_n\geq \gamma'] }{  Q\lefto(\sqrt{2n \snr_n} + Q^{-1}(\alpha_n -\error) \right)  }
\label{eq:beta-beta-awgn-app1-conv}
\end{equation}
where $\gamma'$ satisfies $\prob[S_n \geq \gamma' ] = \alpha_n$.
The denominator in the RHS of~\eqref{eq:beta-beta-awgn-app1-conv} admits the same asymptotic expansion as~\eqref{eq:denominator-beta-beta-awgn}. 
To evaluate the numerator in the RHS of~\eqref{eq:beta-beta-awgn-app1-conv}, we repeat the steps as in~\eqref{eq:app-berry-esseen-awgn}--\eqref{eq:def-gamma_tilde} for $\gamma'$ and obtain
\begin{equation}
\gamma' \geq 2n(1+P_n) +\sqrt{4n(1+2P_n)} Q^{-1}(\alpha_n + k_1/\sqrt{n}) \define \widetilde{\gamma}'.
\label{eq:gamma-p} 
\end{equation}
Furthermore, we have
\begin{IEEEeqnarray}{rCl}
\log \prob[L_n \geq \gamma'] &\leq & \log \prob[L_n \geq\widetilde{\gamma}']\\
&\leq &  -\frac{\log e}{2}\Big\{\widetilde{\gamma}' -\sqrt{n(\widetilde{\gamma}' -n)} \Big\} \label{eq:bound-cdf-chi-squared-c} \IEEEeqnarraynumspace\\
 &=&  - \frac{1}{2} n P_n^2 \log e + \littleo(nP_n^2), \,\, nP_n^2 \to \infty. \IEEEeqnarraynumspace
 \label{eq:bound-cdf-chi-squared-c3}
\end{IEEEeqnarray}
Here,~\eqref{eq:bound-cdf-chi-squared-c} follows from~\cite[Eq.~(4.3)]{laurent2000-05a}, and~\eqref{eq:bound-cdf-chi-squared-c3} follows from~\eqref{eq:gamma-p}  and from algebraic manipulations. 
Finally, we conclude the proof of Theorem~\ref{thm:energy-per-bit}  by substituting~\eqref{eq:denominator-beta-beta-awgn} and~\eqref{eq:bound-cdf-chi-squared-c3} into~\eqref{eq:beta-beta-awgn-app1-conv}.
\end{appendix}

}

\ifthenelse{\boolean{conf}}{
\bibliographystyle{IEEEtran}
\bibliography{IEEEabrv,publishers,confs-jrnls,WeiBib}

\begin{thebibliography}{10}
\providecommand{\url}[1]{#1}
\csname url@samestyle\endcsname
\providecommand{\newblock}{\relax}
\providecommand{\bibinfo}[2]{#2}
\providecommand{\BIBentrySTDinterwordspacing}{\spaceskip=0pt\relax}
\providecommand{\BIBentryALTinterwordstretchfactor}{4}
\providecommand{\BIBentryALTinterwordspacing}{\spaceskip=\fontdimen2\font plus
\BIBentryALTinterwordstretchfactor\fontdimen3\font minus
  \fontdimen4\font\relax}
\providecommand{\BIBforeignlanguage}[2]{{%
\expandafter\ifx\csname l@#1\endcsname\relax
\typeout{** WARNING: IEEEtran.bst: No hyphenation pattern has been}%
\typeout{** loaded for the language `#1'. Using the pattern for}%
\typeout{** the default language instead.}%
\else
\language=\csname l@#1\endcsname
\fi
#2}}
\providecommand{\BIBdecl}{\relax}
\BIBdecl

\bibitem{csiszar11}
I.~Csisz\'{a}r and J.~K\"{o}rner, \emph{Information Theory: Coding Theorems for
  Discrete Memoryless Systems}, 2nd~ed.\hskip 1em plus 0.5em minus 0.4em\relax
  Cambridge, U.K.: Cambridge Univ. Press, 2011.

\bibitem{lapidoth03-10a}
A.~Lapidoth and S.~M. Moser, ``Capacity bounds via duality with applications to
  multiple-antenna systems on flat-fading channels,'' \emph{{IEEE} Trans. Inf.
  Theory}, vol.~49, no.~10, pp. 2426--2467, Oct. 2003.

\bibitem{verdu90-09}
S.~Verd\'{u}, ``On channel capacity per unit cost,'' \emph{{IEEE} Trans. Inf.
  Theory}, vol.~36, no.~5, pp. 1019--1030, Sep. 1990.

\bibitem{arimoto1972-01a}
S.~Arimoto, ``An algorithm for computing the capacity of arbitrary discrete
  memoryless channels,'' \emph{{IEEE} Trans. Inf. Theory}, vol.~18, no.~1, pp.
  14--20, Jan. 1972.

\bibitem{blahut1972-07a}
R.~E. Blahut, ``Computation of channel capacity and rate-distortion function,''
  \emph{{IEEE} Trans. Inf. Theory}, vol.~18, no.~4, pp. 460--473, Jul. 1972.

\bibitem{gallager79-u}
\BIBentryALTinterwordspacing
R.~G. Gallager, ``Source coding with side information and universal coding,''
  1979. [Online]. Available:
  \url{http://web.mit.edu/gallager/www/papers/paper5.pdf}
\BIBentrySTDinterwordspacing

\bibitem{shamai97-03}
S.~{Shamai (Shitz)} and S.~Verd\'{u}, ``The empirical distribution of good
  codes,'' \emph{{IEEE} Trans. Inf. Theory}, vol.~43, no.~3, pp. 836--846, May
  1997.

\bibitem{polyanskiy14-01}
Y.~Polyanskiy and S.~Verd\'{u}, ``Empirical distribution of good channel codes
  with non-vanishing error probability,'' \emph{{IEEE} Trans. Inf. Theory},
  vol.~60, no.~1, pp. 5--21, Jan. 2014.

\bibitem{neyman33a}
J.~Neyman and E.~S. Pearson, ``On the problem of the most efficient tests of
  statistical hypotheses,'' \emph{Philosophical Trans. Royal Soc. A}, vol. 231,
  pp. 289--337, 1933.

\bibitem{feinstein54a}
A.~Feinstein, ``A new basic theorem of information theory,'' \emph{IRE Trans.
  Inform. Theory}, vol.~4, no.~4, pp. 2--22, 1954.

\bibitem{polyanskiy10-05}
Y.~Polyanskiy, H.~V. Poor, and S.~Verd{\'u}, ``Channel coding rate in the
  finite blocklength regime,'' \emph{{IEEE} Trans. Inf. Theory}, vol.~56,
  no.~5, pp. 2307--2359, May 2010.

\bibitem{verdu96-01a}
S.~Verd\'{u}, ``The exponential distribution in information theory,''
  \emph{{Probl. Inf. Transm.}}, vol.~32, no.~1, pp. 86--95, 1996.

\bibitem{riedl2011-10a}
T.~J. Riedl, T.~P. Coleman, and A.~C. Singer, ``Finite block-length achievable
  rates for queuing timing channels,'' in \emph{Proc. IEEE Inf. Theory Workshop
  (ITW)}, Paraty, Oct. 2011, pp. 200--204.

\bibitem{verdu02-06}
S.~Verd\'{u}, ``Spectral efficiency in the wideband regime,'' \emph{{IEEE}
  Trans. Inf. Theory}, vol.~48, no.~6, pp. 1319--1343, Jun. 2002.

\bibitem{polyanskiy11-08a}
Y.~Polyanskiy and S.~Verd\'{u}, ``Scalar coherent fading channel: dispersion
  analysis,'' in \emph{Proc. IEEE Int. Symp. Inf. Theory (ISIT)}, Saint
  Petersburg, Russia, Aug. 2011, pp. 2959--2963.

\bibitem{shannon57}
C.~E. Shannon, ``Certain results in the coding theory for noisy channels,''
  \emph{Inf. Contr.}, vol.~1, pp. 6--25, 1957.

\bibitem{wang2009-07a}
L.~Wang, R.~Colbeck, and R.~Renner, ``Simple channel coding bounds,'' in
  \emph{Proc. IEEE Int. Symp. Inf. Theory (ISIT)}, Seoul, Korea, Jul. 2009.

\bibitem{polyanskiy10-09a}
Y.~Polyanskiy and S.~Verd\'{u}, ``Arimoto channel coding converse and
  {R}\'{e}nyi divergence,'' in \emph{Proc. 48th Allerton Conf. Commun., Contr.,
  Comp.}, Monticello, IL, USA, Sep. 2010.

\bibitem{polyanskiy13}
Y.~Polyanskiy, ``Saddle point in the minimax converse for channel coding,''
  \emph{{IEEE} Trans. Inf. Theory}, vol.~59, no.~5, pp. 2576--2595, May 2013.

\bibitem{erven2014-07a}
T.~{Van Erven} and P.~Harremo\"{e}s, ``R\'{e}nyi divergence and
  {Kullback-Leibler} divergence,'' \emph{{IEEE} Trans. Inf. Theory}, vol.~60,
  no.~7, pp. 3797--3820, Jul. 2014.

\bibitem{gallager68}
R.~G. Gallager, \emph{Information Theory and Reliable Communication}.\hskip 1em
  plus 0.5em minus 0.4em\relax New York, NY, USA: John Wiley \& Sons, 1968.

\bibitem{anantharam1996-01a}
V.~Anantharam and S.~Verd\'{u}, ``Bits through queues,'' \emph{{IEEE} Trans.
  Inf. Theory}, vol.~42, no.~1, pp. 4--18, Jan. 1996.

\bibitem{polyanskiy11-08b}
Y.~Polyanskiy, H.~V. Poor, and S.~Verd\'{u}, ``Minimum energy to send $k$ bits
  through the {G}aussian channel with and without feedback,'' \emph{{IEEE}
  Trans. Inf. Theory}, vol.~57, no.~8, pp. 4880--4902, Aug. 2011.

\bibitem{feller70a}
W.~Feller, \emph{An Introduction to Probability Theory and Its
  Applications}.\hskip 1em plus 0.5em minus 0.4em\relax New York, NY, USA: John
  Wiley \& Sons, 1970, vol.~1.

\bibitem{wu2012-03a}
Y.~Wu and S.~Verd\'{u}, ``Functional properties of minimum mean-square error
  and mutual information,'' \emph{{IEEE} Trans. Inf. Theory}, vol.~58, no.~3,
  pp. 1289--1301, Mar. 2012.

\bibitem{molavianJazi15-12a}
E.~Molavian{J}azi and J.~N. Laneman, ``A second-order achievable rate region
  for {G}aussian multi-access channels via a central limit theorem for
  functions,'' \emph{{IEEE} Trans. Inf. Theory}, vol.~61, no.~12, pp.
  6719--6733, 2015.

\bibitem{iri2015-06a}
N.~Iri and O.~Kosut, ``Third-order coding rate for universal compression of
  {M}arkov sources,'' in \emph{Proc. IEEE Int. Symp. Inf. Theory (ISIT)}, Hong
  Kong, China, Jun. 2015, pp. 1996--2000.

\bibitem{polyanskiy2015-04a}
\BIBentryALTinterwordspacing
Y.~Polyanskiy and Y.~Wu, ``Wasserstein continuity of entropy and outer bounds
  for interference channels,'' Apr. 2015. [Online]. Available:
  \url{http://arxiv.org/abs/1504.04419}
\BIBentrySTDinterwordspacing

\bibitem{verdu98}
S.~Verd\'{u}, \emph{Multiuser Detection}.\hskip 1em plus 0.5em minus
  0.4em\relax Cambridge, UK: Cambridge University Press, 1998.

\bibitem{inglot06-09a}
T.~Inglot and T.~Ledwina, ``Asymptotic optimality of new adaptive test in
  regression model,'' \emph{Ann. Inst. H. Poincar\'{e}}, vol.~42, no.~5, pp.
  579--590, Sep. 2006.

\bibitem{laurent2000-05a}
B.~Laurent and P.~Massart, ``Adaptive estimation of a quadratic functional by
  model selection,'' \emph{Ann. Stat.}, vol.~28, no.~5, pp. 1302--1338, May
  2000.

\end{thebibliography}
}
{

}

\end{document}